\documentclass[usenatbib,usedcolumn,usegraphicx,useAMS]{mn2e}
\usepackage{textcomp}
\usepackage{amsmath}
\usepackage{array}
\usepackage{booktabs}
\begin{document}

\newcommand{\kms}{km\ s$^{-1}$}
\newcommand{\ergs}{erg\ s$^{-1}$} 
\newcommand{\OII}{$\left[O_{\ II}\right]$}
\newcommand{\Deg}{^{\circ}}
\newcommand{\DFK}{D$_{\rm n}$(4000)}
\newcommand{\Hd}{H$\delta$}

\newcommand\aj{AJ} 
\newcommand\araa{ARA\&A} 
\newcommand\apj{ApJ} 
\newcommand\apjl{ApJ} 
\newcommand\apjs{ApJS} 
\newcommand\ao{Appl.~Opt.} 
\newcommand\apss{Ap\&SS} 
\newcommand\aap{A\&A} 
\newcommand\aapr{A\&A~Rev.} 
\newcommand\aaps{A\&AS} 
\newcommand\nat{Nature} 
\newcommand\pasp{PASP} 
\newcommand\pasj{PASJ} 
\newcommand\procspie{Proc.~SPIE} 
\newcommand\memras{MmRAS} 
\newcommand\mnras{MNRAS} 

\title{X-Ray Emitting Active Galactic Nuclei from z = 0.6-1.3 in the Intermediate and High-Density Environments of the ORELSE Survey}

\author[N. Rumbaugh et al.]{
N. Rumbaugh$^1$,
B. C. Lemaux$^1$,
A. Tomczak$^1$,
D. D. Kocevski$^2$,
L. M. Lubin$^1$,\newauthor
P.-F. Wu$^3$,
R. R. Gal$^3$,
L. Shen$^1$,
A. Mansheim$^1$,
C. D. Fassnacht$^1$,
G. K. Squires$^4$\\
$^1$Department of Physics, University of California, Davis, 1 Shields Avenue, Davis CA 95616, USA\\
$^2$Department of Physics and Astronomy, Colby College, Waterville, ME 04901, USA\\
$^3$University of Hawai'i, Institute for Astronomy, 2680 Woodlawn Drive, HI 96822, USA\\
$^4$Spitzer Science Center, California Institute of Technology, M/S 220-6, 1200 E. California Blvd., Pasadena, CA, 91125, USA
}

\maketitle

\begin{abstract}

We studied AGN activity in twelve LSSs in the ORELSE survey, at $0.65<z<1.28$, using a combination of {\it Chandra} observations, optical and NIR imaging and spectroscopy. We located a total of 61 AGNs across our sample that were successfully matched to optical counterparts in the LSSs. Seeking to study AGN triggering mechanisms, we examined the spatial distribution of the AGNs and their average spectral properties. We found that AGN populations across our sample had less time since the last starburst than the overall galaxy populations. We did not find any relation between AGN activity and location within the LSSs, suggesting triggering mechanisms which depend on global environment are at most sub-dominant. To focus on differences between our AGNs, we grouped them into four sub-samples based on the spectral properties of their parents LSSs. We found one of the sub-samples, SG0023 \& SC1604, stood out from the others: AGNs in this sample were disproportionately luminous, their average time since the last starburst event was the smallest, despite the fact that this was not true of the overall galaxy population in those LSSs, and both the AGNs and the overall galaxy population had the largest fraction of close kinematic pairs, which indicates a higher rate of galaxy mergers and interactions. These results suggest that major mergers are driving AGN activity in SG0023 \& SC1604, while other processes are likely triggering less luminous AGNs in the rest of our sample. Additionally, minor mergers are unlikely to play a significant role, since the same conditions that lead to more major mergers should should also lead to more minor mergers, which is not observed in SG0023 \& SC1604. 

\end{abstract}

\begin{keywords}
  Galaxies: active ---
  Galaxies: nuclei ---
  Galaxies: clusters: general ---
  Galaxies: evolution ---
  X-rays: galaxies
\end{keywords}

\section{Introduction}
\label{Intro}

The current consensus is that all massive galaxies contain a supermassive black hole (SMBH) \citep[e.g.,][]{KR95,ford98,graham16}. While the presence of SMBHs can be inferred through their gravitational influences, they can also be directly observed as they accrete material, fueling emission across a wide spectrum. While most SMBHs have low luminosities that preclude observation, a small subset, active galactic nuclei (AGNs), are some of the most powerful emitters in the universe ($L\ga 10^{42}$\ \ergs). These objects are interesting to astronomers and astrophysicists, as numerous studies have found connections between the evolution of the central SMBH and that of its host galaxy. In the local universe, the masses of SMBHs and the masses and velocity dispersions of their hosts' central bulges or spheroids have been shown to be correlated \citep{FM00,trem02,MH03,HR04,xu07,graham16}. As redshift increases up to $z\sim 2$, so do both star formation rates and AGN fractions \citep{BoyTer98,silver08,bluck11}. Across a wide redshift range, AGNs have been observed to be associated with star-forming galaxies \citep[e.g.,][]{kauf03,heck04,vei09,laird10,georg11,jun13,lemaux14Herschel}. 

While the evidence for a connection between SMBH and galaxy evolution is strong, consensus has not yet been reached on the processes that fuel SMBH growth and AGN activity. A number of mechanisms that can trigger AGN activity have been proposed. One such mechanism is major galaxy mergers, which can create inflows of gas to galactic nuclei, fueling bulge growth and an AGN \citep{hopkins06,nar08}. Many others have been proposed, as well, including minor mergers or tidal interactions \citep{moore96,men01,younger08,georg09}, processes such as disk or bar instabilities or turbulence \citep{elm98,genzel08,younger08,HQ10,bour11}, and recycling of stellar material \citep{CO07}.

There is growing evidence that no one mechanism is responsible for creating all AGNs. While major galaxy mergers are thought to fuel some of the brightest AGNs, they are not common, and are likely not responsible for all AGNs \citep[See, e.g., ][]{wild07,reichard09,KH13,heck14}. Some evidence has been found of an association between merging galaxies or those with signs of recent mergers or interactions \citep{koc09b,koss10,trei12,ell13,HI16}, but it is likely this is representative of only a subset of the AGN population, since large numbers of AGNs have been found associated with hosts not likely to have recently undergone major mergers \citep{georg09,cis11,scha11,koc12,rosario12,koul14}. While major mergers are thought to be important at the high-luminosity end, there is a dearth of evidence supporting alternative mechanisms for triggering lower luminosity AGNs, and a multitude of options. 

The question, now, is what are exactly are the dominant AGN triggering mechanisms, and what conditions lead to the different to their activation? Optimally, AGNs should be examined across a wide range of redshifts and environments. We have assembled such a sample as part of the Observations of Redshift Evolution in Large-Scale Environments \citep[ORELSE;][]{lubin09} survey. The ORELSE survey is a systematic search for large-scale structures (LSSs) around an original sample of 20 galaxy clusters in a redshift range of $0.6 < z < 1.3$, the over-arching goal being to study galaxy properties over a wide range of local and global environments. The survey has revealed superclusters and merging systems, and has also found some of the initially targeted galaxy clusters to be isolated systems. This provides an excellent laboratory for studying AGN activity. The wide range of environments means the sample likely contains infalling populations where major mergers between blue galaxies are more prevalent, and one system, the Cl1604 supercluster, has already been shown to have AGNs associated with mergers \citep{koc09b}. Additionally, both star formation and AGN activity are known to increase with redshift \citep{BoyTer98,silver08,bluck11}, so the higher redshift of our sample, and the wide range, provide an ideal set for examining AGN activity and how it relates to galaxy evolution.

Twelve of the systems in the ORELSE survey have {\it Chandra} imaging with which we can study AGNs at the redshifts of the LSSs. We have extensive optical imaging and spectroscopy across the sample to complement the X-ray dataset. In this paper, we use the the {\it Chandra} imaging of these twelve systems to locate AGNs within them and measure their X-ray luminosities. We couple this with the optical properties of their hosts, and the underlying galaxy populations, to investigate AGN triggering mechanisms and their relation to galaxy evolution. For our cosmological model, we assume $\Omega_m=0.3$, $\Omega_{\Lambda}=0.7$, and $h = H_0/70 km s^{-1} Mpc^{-1}$. 

We first discuss the LSSs in our sample in Section \ref{sec:samp}. We describe our observations, data reduction, and methods for locating AGNs in Section \ref{sec:red}. In Section \ref{sec:global}, we examine the properties of the global galaxy populations in our sample, and divide our sample into several sub-samples based on their properties. We then analyze the AGN populations in Section \ref{sec:AGN}, followed by a discussion in Section \ref{sec:dis}.

\section{The ORELSE Large-scale Structure Sample}
\label{sec:samp}

In this section we describe our sample, which
is more succinctly summarized in Table \ref{strsumtab}. As described in Section \ref{Intro}, our sample consists of all LSSs in the ORELSE survey with useful {\it Chandra} imaging. This consists of twelve systems\footnote{A thirteenth system, Cl1137+3000, also has {\it Chandra} imaging, but the shallow exposure (less than half the next shortest exposure time in our sample) and the high redshift of the cluster mean that it is not useful for our analysis.}, composed of both observations taken specifically for the ORELSE survey and other publicly available archived data. 

The redshift boundaries defined below were determined by visually
examining each LSS's spectroscopic redshift histogram. Delineating where
LSSs end is not straightforward, complicated by associated
filaments or possible nearby sheets. The redshift boundaries are
chosen conservatively, in the sense of including all galaxies which
could be part of each overall large-scale structure.

For descriptions of the Cl0023 supergroup, the RXJ0910 LSS, the Cl1324 supercluster, the Cl1604 supercluster, and the clusters RXJ1757.3+6631
 and RXJ1821.6+6827 (hereafter SG0023, RXJ0910, SC1324, SC1604, RXJ1757, and RXJ1821, respectively), see \citet{rum12}, \citet{rumb13}, and the references therein. 

\begin{table*}
\caption{Properties of Observed ORELSE LSSs}
\label{strsumtab}
\begin{tabular}{llllllrrllr}
\toprule
\footnotesize{LSS}
 & \footnotesize{R.A.$^a$}
 & \footnotesize{Dec.$^a$}
 & \footnotesize{$\langle z\rangle$}
 & \footnotesize{$z$ Lower}
 & \footnotesize{$z$ Upper}
 & \footnotesize{Num. of}
 & \footnotesize{Num. of}
 & \footnotesize{$\sigma$}
 & \footnotesize{Confirmed} 
 & \footnotesize{Confirmed} \\
   { }
 & \footnotesize{(J2000)}
 & \footnotesize{(J2000)}
 & { }
 & \footnotesize{Bound}
 & \footnotesize{Bound}
 & \footnotesize{Known}
 & \footnotesize{Known}
 & \footnotesize{Range$^b$}
 & \footnotesize{Members$^c$}
 & \footnotesize{AGN$^c$} \\
   { }
 & { }
 & { }
 & { }
 & { }
 & { }
 & \footnotesize{Clusters}
 & \footnotesize{Groups}
 & { }
 & { }
 & { }\\
\midrule
SG0023 & 00\ 23\ 51 & +04\ 22\ 55 & 0.84 & 0.82 & 0.87 & 0 & 5 & 200-500  & 244 & 7 \\
RCS0224 & 02\ 24\ 36 & -01\ 55\ 58 & 0.77 & 0.76 & 0.79 & 2 & 1 & 200-800 & 119 & 4 \\
Cl0849 & 08\ 48\ 47 & +44\ 54\ 06 & 1.26 & 1.25 & 1.28 & 2 & 0 & $^d$ & 74 & 4 \\
RXJ0910 & 09\ 10\ 40 & +54\ 19\ 57 & 1.11 & 1.08 & 1.15 & 2 & 0 & 500-900 & 142 & 9 \\
RXJ1053 & 10\ 53\ 40.2 & +57\ 35\ 22.3 & 1.14 & 1.10 & 1.15 & 1-2 & 0 & $880\pm130$ & 72 & 1 \\
RXJ1221 & 12\ 21\ 26.1 & +49\ 18\ 30.7 & 0.70 & 0.69 & 0.71 & 1 & 1 & 800-850 & 160 & 4 \\
SC1324 & 13\ 24\ 45 & +30\ 34\ 18 & 0.76 & 0.65 & 0.79 & 3 & 1 & 150-1000 & 454 & 8 \\
Cl1350 & 13\ 50\ 48.3 & +60\ 07\ 11.5 & 0.80 & 0.79 & 0.81 & 1 & 2 & 200-1000 & 102 & 3 \\
SC1604 & 16\ 04\ 15 & +43\ 16\ 24 & 0.90 & 0.84 & 0.96 & 5 & 3 & 150-1100 & 531 & 10 \\
RXJ1716 & 17\ 16\ 54 & +67\ 08\ 47 & 0.81 & 0.80 & 0.83 & 2 & 0 & 750-1150 & 144 & 4 \\
RXJ1757 & 17\ 57\ 19.0 & +66\ 31\ 27.8 & 0.69 & 0.68 & 0.71 & 1 & 0 & $540\pm140$  & 75 & 2 \\
RXJ1821 & 18\ 21\ 32.3 & +68\ 27\ 55.4 & 0.82 & 0.80 & 0.84 & 1 & 0 & $1150\pm120$  & 131 & 5 \\
\bottomrule
\multicolumn{11}{p{17cm}}{$^{\rm a}$ \footnotesize{Coordinates for LSSs with more than one cluster are the approximate central positions, while the others are given as the centroid of the peak of diffuse X-ray emission associated with the respective cluster.}}\\
\multicolumn{11}{p{17cm}}{$^{\rm b}$ \footnotesize{In units of \kms. For LSSs with more than group or cluster, this measurement is the range of velocity dispersions of groups and clusters within the LSS.  All velocity dispersions are measured within 1 Mpc.}}\\
\multicolumn{11}{p{17cm}}{$^{\rm c}$ \footnotesize{Spectroscopically confirmed objects ($Q=3$,4) within the redshift bounds of the LSS; see Section \ref{sec:optobs} for quality flag details.}}\\
\multicolumn{11}{l}{$^{\rm d}$ \footnotesize{Velocity dispersion not measured.}}
\end{tabular}
\end{table*}

\subsection{RCS J0224-0002.5}

The optically-selected $z = 0.77$ galaxy cluster RCS J0224-0002.5 (hereafter RCS0224) was discovered by \citet{glad02} as part of the Red-Sequence Cluster Survey \citep[RCS; ][]{GY05}. They found two to four lensed background objects, and fit a lensing model consistent with a velocity dispersion of $\sim1000$ \kms for the cluster. \citet{hicks07} used 101 ks of {\it Chandra} observations to measure an integrated cluster temperature of $5.1^{+0.9}_{-0.5}$ keV. They found the cluster to be in agreement with several scaling relations for virialized clusters. Using the same data, \citet{hicks08} measured an X-ray luminosity within $R_{500}$ of $4.4\pm0.5 \times 10^{44}$ ergs s$^{-1}$. 

\subsection{Cl0849}

The Cl0849 LSS contains at least two clusters within close proximity and at known similar redshifts of $z \sim 1.26$. The IR-selected galaxy cluster RX J0848.6+4453 at $z=1.26$ was discovered as part of a deep field survey \citep{stan97}. A soft-band (0.5-2.0 keV) X-ray luminosity was measured of $0.8\pm0.3\times 10^{44}$ ergs s$^{-1}$ using archival {\it ROSAT} data \citep{ros99}. Only $4\farcm2$ away, RX J0848.9+4452 was discovered as part of the {\it ROSAT} Deep Cluster Survey \citep{ros99}. They measured a soft-band X-ray luminosity of $1.5\pm0.3\times 10^{44}$ ergs s$^{-1}$. {\it Chandra} imaging of RX J0848.9+4452 yielded an X-ray temperature of 5.8$^{+2.8}_{-1.7}$ keV and a bolometric X-ray luminosity of 3.3$^{+0.9}_{-0.5} \times 10^{44}$ ergs s$^{-1}$ \citep{stan01}, in approximate agreement with scaling relations. 

\subsection{RXJ1053+5735}

RX J1053+5735 (hereafter RXJ1053), an X-ray selected cluster at $z=1.14$, was discovered in {\it ROSAT} ultra-deep HRI imaging of the Lockman Hole \citep{has98}. They observed an unusual double-lobed profile in the extended X-ray emission. Using XMM-Newton, \citet{hash02} measured an X-ray temperature of 4.9$^{+1.5}_{-0.9}$ keV and a bolometric X-ray luminosity of $3.4\pm0.34\times 10^{44}$ ergs s$^{-1}$. Both X-ray redshifts and redshifts of member galaxies are concordant between the lobes, which are only $\sim$ 250 kpc apart, suggesting an equal-mass cluster merger \citep{hash05}. 

\subsection{RXJ1221+4918}

RXJ1221+4918 (hereafter RXJ1221) is an X-ray-selected cluster at $z=0.70$. It was serendipitously detected in {\it ROSAT} observations for the 160 deg$^2$ survey \citep{vik98}. \citet{vik02} measured an X-ray temperature of 7.2$\pm$0.6 keV for the cluster and soft-band and bolometric X-ray luminosities of 7.0 and 28.7 $\times 10^{44}$ ergs s$^{-1}$, respectively. There has been repeated analysis of the cluster, with  \citet{mullis03} measuring a soft-band X-ray luminosity of 4.27 $\times 10^{44}$ ergs s$^{-1}$ for the cluster, while \citet{ett04} measured an X-ray temperature of 7.5$^{+0.7}_{-0.6}$ keV and a bolometric X-ray luminosity of $12.95\pm0.39\times 10^{44}$ ergs s$^{-1}$, and \citet{vik09b} measured a temperature and soft-band luminosity of 6.63$\pm$0.75 keV and $3.35\times 10^{44}$ ergs s$^{-1}$, respectively.

\subsection{Cl1350.0+6007}

Cl1350.0+6007 (hereafter Cl1350) is an X-ray selected cluster at $z=0.80$, discovered in the {\it ROSAT} Deep Cluster Survey \citep[RDCS; ][]{ros98}. Using {\it Chandra} data, \citet{holden02} measured an X-ray temperature of 4.9$^{+0.7}_{-0.6}$ keV and a bolometric X-ray luminosity of $4.1^{+0.5}_{-0.4}\times 10^{44}$ ergs s$^{-1}$. Also with {\it Chandra}, \citet{jel05} measured an X-ray luminosity of $2.3\times 10^{44}$ ergs s$^{-1}$ in the 0.3-7.0 keV range.

\subsection{RXJ1716.9+6708}

RXJ1716.9+6708 (hereafter RXJ1716) is an X-ray selected cluster at $z=0.81$. \citet{henry97} discovered the cluster in the {\it ROSAT} All-Sky Survey \citep{voges96} of the North Ecliptic Pole, and measured a soft-band X-ray luminosity of $3.2\pm0.9\times 10^{44}$ ergs s$^{-1}$. Follow-up observations using the {\it ROSAT} High Resolution Imager yielded an X-ray temperature of 5.66$^{+1.37}_{-0.58}$ keV and a hard 2-10 keV X-ray luminosity of $8.19\pm 0.43\times 10^{44}$ ergs s$^{-1}$ in the 0.3-7.0 keV range. Using {\it Chandra} data, \citet{jel05} measured an X-ray luminosity of $5.6\times 10^{44}$ ergs s$^{-1}$ in the 0.3-7.0 keV range. See Lemaux et al. (2016, in prep.) for more details on this LSS. 

\section{Observations and Reduction}
\label{sec:red}
\subsection{Optical and NIR Observations}
\label{sec:optobs}

In analyzing the AGN hosts, we use ground-based imaging data obtained for each field with either the Large Format Camera \citep[LFC;][]{simcoe00} on the Palomar 5m telescope or the Suprime-Cam \citep{Suprime-Cam} on the Subaru 8-m telescope. The LFC observations were taken using Sloan Digital Sky Survey (SDSS)-like $r'$, $i'$, and $z'$ filters. We hereafter refer to these LFC filters, or the equivalents from the Suprime-Cam data, as R, I, and Z. Additionally, photometric redshift catalogs have been created for a subset of our sample, which we use to examine the effects of completeness on our analysis and to calibrate our `supercolor' parameterization described in Section \ref{TRSATBP}. For these catalogs, we use the LFC and Subaru data (in total, B, V, Rc, R+, Ic, I+, Z+, and Y band data were taken using Suprime-Cam), as well as J and K band data taken using the United Kingdom Infrared Telescope Wide Field Camera \citep[WFCAM;][]{WFCAM}, J and Ks band data using the Wide-field InfraRed Camera \citep[WIRCam;][]{WIRCam} on the Canada-France-Hawai'i Telescope, and 3.6, 4.5, 5.8, and 8.0 micron data using the Infrared Array Camera \citep[IRAC;][]{IRAC} on the Spitzer telescope. Which telescope data are available varies by LSS. For more descriptions of the reductions, see \citet{gal08}, Lemaux et al. (2016, in prep.), and Tomczak et al. (2016, in prep.). 

SC1604 was also imaged using ACS. The HST imaging for SC1604 consists of 17 ACS pointings designed to image nine of the ten galaxy density peaks in the field. Observations were taken using the F606W and F814W bands. These bands roughly correspond to broadband V and I, respectively. See \citet{koc09b} for details on the ACS reduction.

\subsubsection{Photometric Redshift Catalogs}
\label{sec:photoz}

To extract photometric redshifts, and other characteristics of our galaxy populations, we performed spectral energy distribution (SED) fitting on our imaging data. Results from aperture photometry were used to run the Easy and Accurate $z_{phot}$ from Yale \citep[EAZY;][]{EAZY} code, which performs an iterative $\chi^2$ fit using Projet d'\'{E}tude des GAlaxies par Synth\`{e}se \'{E}volutive \citep[P\'{E}GASE;][]{PEGASE} models. This code outputs $P\left(z\right)$, a measure of our confidence that the respective source is at a given redshift. The redshift at which $P\left(z\right)$ peaks was adopted as the photometric redshift, $z_{peak}$. A second round of fitting with purely stellar templates was carried out to locate the stars in our sample. We cut likely stars and objects with a S/N $< 3$ in the detection band, those covered in less than five of the broadband images, with significant saturation, or with poor fits to galaxy SEDs. 

A final stage of fitting was carried out using the Fitting and Assessment of Synthetic Templates \citep[FAST;][]{FAST} code. High quality spectroscopic redshifts were used when available, and $z_{peak}$ from EAZY was used as a redshift prior for all other cases. From the EAZY output, we derived rest-frame magnitudes for all our sources, except those excised as described above. This SED fitting has so far been carried out on SG0023, RXJ0910, SC1324, SC1604, RXJ1716, RXJ1757, and RXJ1821. See Lemaux et al. (2016, in prep.) and Tomczak et al. (2016, in prep.) for more details on the photometric catalogs and SED fitting. 

\subsubsection{Spectroscopy}

Our photometric catalog is complemented by extensive spectroscopic data. For this part of the study, we used the Deep Imaging Multi-Object Spectrograph \citep[DEIMOS;][]{faber03} on the Keck II 10m telescope. In addition, SG0023, RXJ0910, SC1604 and RXJ1821 have some LRIS coverage \citep[see][]{oke98,GalLub04,gioia04,tan08}. DEIMOS has a wide field of view ($16\farcm9 \times 5\farcm0$), high efficiency, and is able to position over 120 targets per slit mask, which makes the instrument ideal for establishing an extensive spectroscopic catalog. We targeted objects as faint as $\sim 25$, with most objects having $m_{i'}\le24.5$. On DEIMOS, we used the 1200 line mm$^{-1}$ grating, blazed at 7500 \AA, and 1$\arcsec$-wide slits. These specifications create a pixel scale of 0.33 \AA\ pixel$^{-1}$ and a FWHM resolution of $\sim1.7$\AA. The central wavelength was varied from LSS to LSS and sometimes between different masks for the same LSS. Our setup provided wavelength coverage approximately within 1300\ \AA of the central wavelengths. Central wavelengths for the spectroscopic observations for the twelve LSSs and the approximate spectral coverages are displayed in Table \ref{speccovtab}. When more than one central wavelength was used per LSS, a range is given. Total exposure times for the observations are in the range of 1-4.5 hours per mask and varied based on conditions and $i$'-band distributions of targets.

\begin{table*}
\caption{DEIMOS Spectroscopic Observation Characteristics}
\label{speccovtab}
\begin{tabular}{lllll}
\footnotesize{LSS}
 & \footnotesize{Central}
 & \footnotesize{ Approx. Spectral}
 & \footnotesize{Exp. Time}
 & \footnotesize{Avg. Seeing}\\
 \footnotesize{ }
& \footnotesize{$\lambda$ (\AA) }
& \footnotesize{Coverage (\AA)}
& \footnotesize{Range (s) }
& \footnotesize{Range ($\arcsec$)}\\
\midrule
SG0023 & 7500-7850 & 6200-9150 & 5700-9407 & 0.45-0.81\\
RCS0224 & 7300-7450 & 6000-8750 & 6840-7520 & 0.53-0.88\\
Cl0849 & 8700 & 7400-10000 & 6300-16200 & 0.51-1.5\\
RXJ0910 & 8000-8100 & 6700-9400 & 7200-11664 & 0.5-1.05\\
RXJ1053 & 8200 & 6900-9500 & 7200-9000 & 0.56-0.79\\
RXJ1221 & 7200 & 5900-8500 & 4860-8400 & 0.55-1.2\\
SC1324 & 7200 & 5900-8500 & 2700-10800 & 0.44-1.0\\
Cl1350 & 7500 & 6200-8800 & 3600-10400 & 0.5-1.55\\
SC1604 & 7700 & 6385-9015 & 3600-14400 & 0.5-1.3\\
RXJ1716 & 7800 & 6500-9100 & 5400-9000 & 0.54-0.83\\
RXJ1757 & 7000-7100 & 5700-8400 & 6300-14730 & 0.47-0.82\\
RXJ1821 & 7500-7800 & 6200-9100 & 7200-9000 & 0.58-0.86\\
\bottomrule
\end{tabular}
\end{table*}

Spectroscopic targets were chosen based on color and magnitude, following the method of \citet{lubin09}. The number of spectroscopic targets in each LSS are shown in Table \ref{srcsum}. Redshifts were determined or measured for all targets and given a quality flag value, $Q$, where $Q = 1$ indicates that we could not determine a secure redshift, $Q = 2$ means a redshift was obtained using features that were only marginally detected, $Q = 3$ means one secure and one marginal feature were used to calculate the redshift, and $Q = 4$ meant at least two secure features were used. Those sources determined to be stars were given a flag of $-1$. See \citet{gal08} and \citet{new13} for more details on quality flags and the spectral targeting method. For our analysis, extragalactic redshifts with $Q = 3$,4 were deemed satisfactory, and the number of such sources in each LSS is shown in Table \ref{srcsum}. We measured an average of $\sim$ 800 high-quality redshifts per LSS, with the number of $Q = 3$,4 redshifts in the field of view for each LSS ranging from 410 for RXJ1053 to 1849 for SC1604.

\subsection{X-ray Observations}

All X-ray imaging of the clusters was conducted with the Advanced CCD Imaging Spectrometer (ACIS) of the {\it Chandra} X-ray Observatory. The instrument has 10 CCD's; four are arranged in a square with a $16\farcm9 \times 16\farcm9$ field of view, while the others are arranged in a line parallel to the bottom of this square, with a field of view of $8\farcm3 \times 50\farcm6$. The ACIS-I array has an aimpoint on the I3 chip, near the center of the square area, while the aimpoint of the ACIS-S array is on the S3 chip, near the center in the line array. While most (19) of the observations used for this study employ the ACIS-I array, three observations taken from the archive used the ACIS-S array instead. In either case, the 5-6 CCD's closest to the aimpoint are typically used for each observation. SC1604 and SC1324, with angular sizes in excess of 20$\arcmin$, were observed with two pointings each of the ACIS-I array. For SC1604, the two pointings are meant to cover as much of the LSS as possible, and there is only a small overlap between the two pointings ($\sim 30$ arcminutes$^2$). For SC1324, the two pointings are centered near the two largest and originally discovered clusters, SC1324+3011 and SC1324+3059. There is an approximately 13$\arcmin$ gap between the north and south pointings. All other LSSs were observed with one pointing. Total exposure times ranged from 45-190 ks. Details of the observations are summarized in Table \ref{obsIDtab}.


\begin{table*}
\caption{{\it Chandra} Observations}
\label{obsIDtab}
\begin{tabular}{lllllll}
\toprule
\footnotesize{Observation}
 & \footnotesize{Target}
 & \footnotesize{Instrument}
 & \footnotesize{PI}
 & \footnotesize{Exposure}
 & \footnotesize{RA$^a$}
 & \footnotesize{Dec.$^a$} \\
   \footnotesize{ID}
 & \footnotesize{ }
 & \footnotesize{ }
 & \footnotesize{ }
 & \footnotesize{Time (ks)}
 & \footnotesize{ }
 & \footnotesize{ }\\
\midrule
7914 & SG0023 & ACIS-I & Lubin & 49.38 & 00\ 23\ 52.30 & 04\ 22\ 34.20 \\
3181 & RCS0224 & ACIS-S & Gladders & 14.37 & 02\ 24\ 34.10 & -00\ 02\ 30.90 \\
4987 & RCS0224 & ACIS-S & Ellingson & 88.97 & 02\ 24\ 34.10 & -00\ 02\ 30.90 \\
927 & Cl0849 & ACIS-I & Stanford & 125.15 & 08\ 48\ 55.90 & 44\ 54\ 50.00 \\
1708 & Cl0849 & ACIS-I & Stanford & 61.47 & 08\ 48\ 55.90 & 44\ 54\ 50.00 \\
2227 & RXJ0910 & ACIS-I & Stanford & 105.74 & 09\ 10\ 45.41 & 54\ 22\ 05.00 \\
2452 & RXJ0910 & ACIS-I & Stanford & 65.31 & 09\ 10\ 45.41 & 54\ 22\ 05.00 \\
4936 & RXJ1053 & ACIS-S & Predehl & 92.4 & 10\ 53\ 43.00 & 57\ 35\ 00.00 \\
1662 & RXJ1221 & ACIS-I & van Speybroeck & 79.08 & 12\ 21\ 24.50 & 49\ 18\ 14.40 \\
9403 & SC1324 & ACIS-I & Lubin & 26.94 & 13\ 24\ 49.50 & 30\ 51\ 34.10 \\
9404 & SC1324 & ACIS-I & Lubin & 30.4 & 13\ 24\ 42.50 & 30\ 16\ 30.00 \\
9836 & SC1324 & ACIS-I & Lubin & 20 & 13\ 24\ 42.50 & 30\ 16\ 30.00 \\
9840 & SC1324 & ACIS-I & Lubin & 21.45 & 13\ 24\ 49.50 & 30\ 51\ 34.10 \\
2229 & Cl1350 & ACIS-I & Stanford & 58.31 & 13\ 50\ 46.10 & 60\ 07\ 09.00 \\
6932 & SC1604 & ACIS-I & Lubin & 49.48 & 16\ 04\ 19.50 & 43\ 10\ 31.00 \\
6933 & SC1604 & ACIS-I & Lubin & 26.69 & 16\ 04\ 12.00 & 43\ 22\ 35.40 \\
7343 & SC1604 & ACIS-I & Lubin & 19.41 & 16\ 04\ 12.00 & 43\ 22\ 35.40 \\
548 & RXJ1716 & ACIS-I & van Speybroeck & 51.73 & 17\ 16\ 52.30 & 67\ 08\ 31.20 \\
10443 & RXJ1757 & ACIS-I & Lubin & 21.75 & 17\ 57\ 19.80 & 66\ 31\ 39.00 \\
11999 & RXJ1757 & ACIS-I & Lubin & 24.7 & 17\ 57\ 19.80 & 66\ 31\ 39.00 \\
10444 & RXJ1821 & ACIS-I & Lubin & 22.24 & 18\ 21\ 38.10 & 68\ 27\ 52.00 \\
10924 & RXJ1821 & ACIS-I & Lubin & 27.31 & 18\ 21\ 38.10 & 68\ 27\ 52.00 \\
\bottomrule
\multicolumn{7}{l}{$^{\rm a}$ \footnotesize{Coordinates refer to those of the observation aimpoint.}}
\end{tabular}
\end{table*}

\subsection{X-ray Data Reduction and Photometry}
\label{sec:datared}

The reduction of the data was conducted using the {\it Chandra} Interactive Analysis of Observations 4.7 software \citep[CIAO;][]{frusc06}. We used the Imperial reduction pipeline, which is described in detail in \citet{laird09} and \citet{nandra15}. We will briefly summarize it here. 

For each individual observation, we corrected for aspect offsets, applied a destreaking algorithm, removed bad pixels and cosmic rays, corrected for charge-transfer inefficiency and gain effects, and applied the ACIS particle background cleaning algorithm. Unlike in \citet{nandra15}, data for all available ACIS chips were reduced, instead of restricting the reduction to just ACIS Chips 0-3. This was necessary, as some observations had aim points on other ACIS chips. For each observation, flaring events were detected and excised. After all of these corrections, the astrometry was corrected using our spectroscopic catalogs for reference. 

Event files were then created in four different energy bands: soft (0.5-2.0 keV), hard (2.0-7.0 keV), ultrahard (4.0-7.0 keV), and full (0.5-7.0 keV). Exposure maps were created for each band using the task {\it merge\_obs}, with spectral weights corresponding to a power law with $\gamma = 1.4$. For LSSs with more than one observation, the individual images and exposure maps were then stacked. 

Point source detection was carried out in a two-step process. A `prewavdetect' run came first, where the task {\it wavdetect} was run for each image (in each band) at a probability threshold of 10$^{-4}$, designed to detect virtually all real sources, but also introducing many spurious detections. Aperture photometry was then carried out on all detections in the following manner. Source counts were measured in a circular aperture centered on the detection position with a radius such that 70\% of the source's flux should be enclosed\footnote{The {\it Chandra} point spread function was calculated using the MARX simulator \citep{wise03}. See \citet{laird09}.}. The background was measured in an annulus with an inner radius equal to 1.5 times that of a circle containing 95\% of the source flux, and an outer radius 100 pixels larger. Other point sources were excluded when measuring background counts. The probability of false detections was then calculated using Poisson statistics. Then, the second and final detection iteration was carried out, using a probability threshold of $4\times 10^{-6}$. The `prewavdetect' step corrects for underestimation of the background from masking false positive detections. 

Sources from the second {\it wavdetect} run, in all bands, were merged into a single catalog, and each detection was checked by eye. For our final aperture photometry, we measured soft band fluxes using the 95\% enclosed energy radii, while we used the 90\% enclosed energy radii for the hard and full bands. This is because of the excessively large size of the 95\% enclosed energy radii in these bands. Fluxes were estimated using the Bayesian method described in \citet{laird09}, using a power law spectral model with $\gamma = 1.4$. The aperture photometry was also used to estimate detection significances according to the formula \begin{equation}
\sigma=C/\left(1 + \sqrt{0.75+B}\right)
\label{eq:sig}
\end{equation}
where $C$ and $B$ are the net source counts and background counts within the aperture, respectively.

\subsection{Optical Matching}
\label{sec:OM}

To search for AGNs within the individual clusters, we matched our X-ray point sources to
optical catalogs. Only objects with detection significances, in at least one band, above 2$\sigma$ were considered\footnote{We used 2$\sigma$ as our threshold instead of 3$\sigma$ because being successfully matched to an optical source increases confidence in the detection of an X-ray object.}. We used the
maximum likelihood ratio technique described in \cite{rum12}, which
was developed by \citet{SS92} and also used by \citet{taylor05}, 
\citet{gil07}, and \citet{koc09b}. The main statistic calculated in each case is the
likelihood ratio (LR), which estimates the probability that a given
optical source is the genuine match to a given X-ray point source relative
to the arrangement of the two sources arising by chance. The LR is
given by the equation
\begin{equation}
  LR_{i,j} = \frac{w_i \exp(-r_{i,j}^2/2\sigma_j^2)}{\sigma_j^2}
\end{equation}
Here, $r_{i,j}$ is the separation between objects $i$ and $j$, $\sigma_j$ is the positional error\footnote{Positional errors of X-ray sources were calculated using the method of \citet{kim07}. Optical positional errors are assumed to be negligible in comparison.} of object $j$, and $w_i^2 = 1/n\left(<{m}_I\right)$ is the inverse of the number density of optical sources with magnitude fainter than ${m}_I$. The inclusion of the latter quantity is designed to weight against matching to fainter optical objects that are more likely to have chance projections. For each X-ray source, we carried out a Monte Carlo (MC) simulation to estimate the probability that each optical counterpart is the true match using the LRs. See \citet{rum12} for more details on the method. 

For each LSS, except SC1604, $n\left(<{m}_I\right)$ was measured using $I$ magnitudes from our LFC/Suprime-Cam catalogs. For SC1604, ACS data were also available, but these observations did not cover the entire LSS. All objects were matched to the LFC/Suprime-Cam catalogs. When possible, objects were also matched using the F814W magnitude from the ACS catalogs and matches to ACS objects took precedence. 

\begin{figure*}
\includegraphics[width=0.9\textwidth]{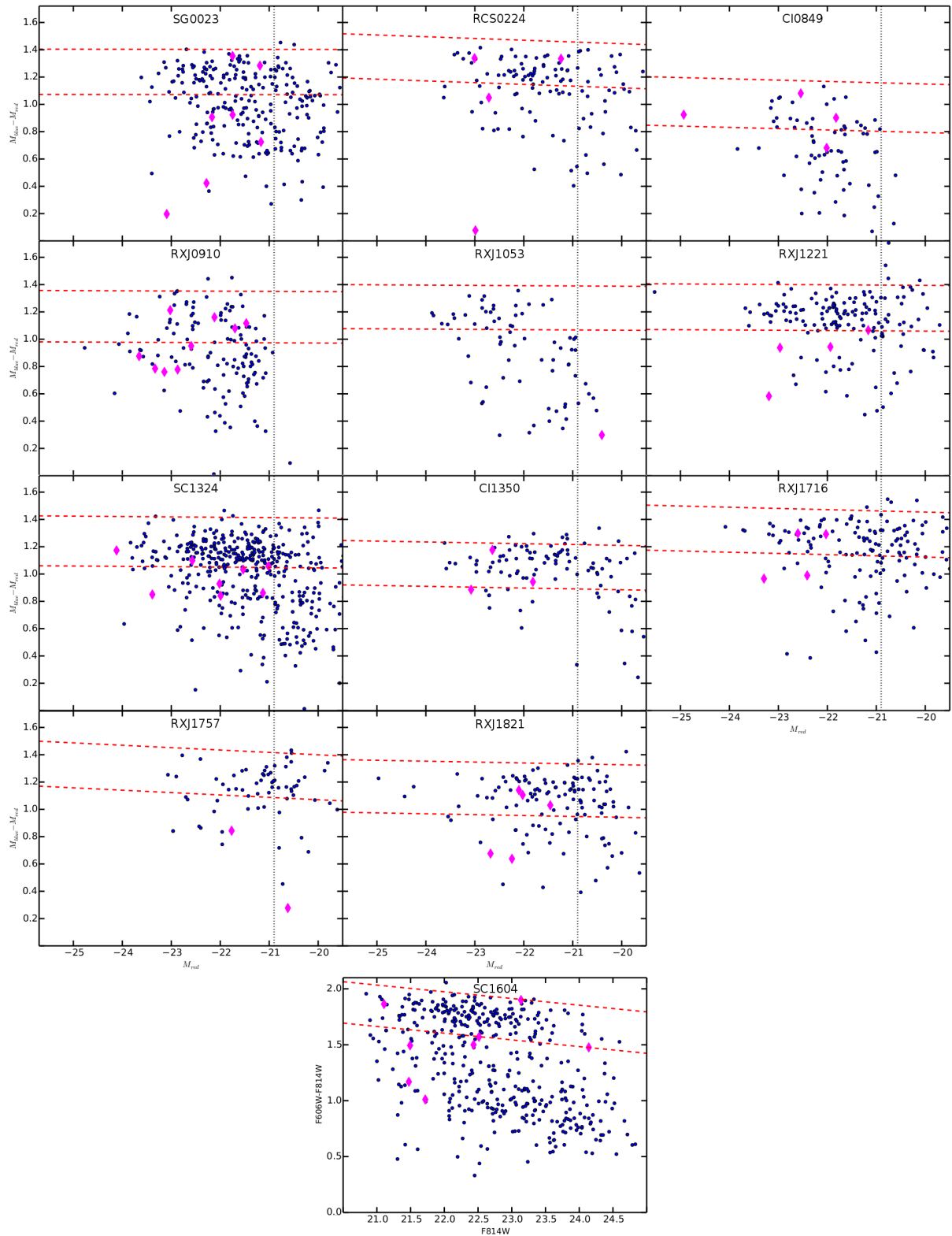}
\caption{
Color-magnitude diagrams for all LSSs are shown. AGN hosts are shown with magenta diamonds, while all other spectroscopically confirmed LSS members are shown with blue circles. Red dashed lines indicate the extent of the red sequence (See Section \ref{TRSATBP} for more details). Vertical dotted lines indicate the magnitude cutoff used for constructing composite spectra described in Section \ref{sec:globspec}. No such line is shown for SC1604, because ACS data were available, and so those colors were used to construct the diagrams, although the parameterized colors were used for the magnitude cutoffs for consistency with the other LSSs. For all other LSSs, parameterizations of the $R$, $I$, and $Z$ bands were used. See Section \ref{TRSATBP} for more details on these parameterizations.
}
\label{CMDS}
\end{figure*}

\section{Global LSS Properties}
\label{sec:global}

The twelve LSSs in our sample span a wide range of redshifts, sizes, and evolutionary states. There are a number of isolated clusters, such as RXJ1757 and RXJ1821, several LSSs in the process of merging, such as the supergroup SG0023 and the merging clusters of RXJ1053, as well as several superclusters, including SC1324 and SC1604. We would like to investigate the differences between the AGNs across this wide range of environments, through which we can learn about their triggering mechanisms and the effects the environments have on active galaxies. To accomplish this, we first study the global properties of the LSSs in our sample.

\subsection{The Red Sequence and the Blue Populations} 
\label{TRSATBP}

A color-magnitude diagram (CMD) provides information on the evolutionary states of a LSS's galaxy population. When the colors of galaxies at a common redshift are plotted against their magnitudes on a CMD, they generally separate out into two regions: the red sequence, which largely contains older, quiescent galaxies, and the blue cloud, which is characterized by younger, actively star-forming populations. Intermediate to these is the green valley, which is usually sparsely populated, and may be a transitional region where blue galaxies are rapidly evolving onto the red sequence after star formation ceases \citep{faber07}, or rejuvenated red sequence galaxies \citep{dress13}. 

Because of the redshift range of our sample, using the same bands for magnitudes and colors across all LSSs would not produce accurate representations of the red sequence and blue cloud. The rest-frame energy range probed by the $R$ band at our lowest redshifts, for example is approximately the same as that probed by the $I$ band at the highest redshifts. So, while an $R-I$\ versus $I$ CMD would be appropriate for the lower redshifts, we would need to use an $I-Z$\ versus $Z$ CMD at higher redshifts. 

Ideally, SED fits for each source could be used to estimate rest-frame magnitudes comparable across different redshifts. As described in Section \ref{sec:optobs}, we have carried out such fitting for a subset of our sample. However, without rest-frame magnitudes for all sources, we cannot use this method. As a solution to this problem, we created a parameterization of the observed optical magnitude that we could use for our entire sample. We sought a parameterization of the form:\begin{align*}
f_{blue}&=A_{blue}\left[1-B_{blue}\left(z-z_0\right)\right]f_{\nu,R}\\&+\left(1-A_{blue}\right)\left[B_{blue}\left(z-z_0\right)\right]f_{\nu,I}\\
f_{red}&=A_{red}\left[1-B_{red}\left(z-z_0\right)\right]f_{\nu,I}\\&+\left(1-A_{red}\right)\left[B_{red}\left(z-z_0\right)\right]f_{\nu,Z}
\end{align*}
To represent the range of galaxies present in our sample, we used four EAZY P\'{E}GASE models: three post-starburst models with ages of 0.1, 0.4, and 5 Gyr, and one constant star-forming model 50 Myr after the inception of star formation with $A_v=3$. We fit our parameterization to minimize the differences in color between our supercolors and the true rest-frame colors for the SEDs across the redshift range of our sample. We found \begin{align*}
f_{blue}&=0.45\left[1-1.824\left(z-0.679\right)\right]f_{\nu,R}\\&+0.55\left[1.824\left(z-0.679\right)\right]f_{\nu,I}\\
f_{red}&=0.424\left[1-1.794\left(z-0.628\right)\right]f_{\nu,I}\\&+0.576\left[1.794\left(z-0.628\right)\right]f_{\nu,Z}
\label{eq:supercolor}
\end{align*}
to be the best fit. From these equations, we calculated the parameterized absolute magnitudes $M_{blue}$ and $M_{red}$, which we will refer to as the supercolors. These can be thought of as $M_{u*}$ and $M_{B}$, respectively. For our color analyses, we use these supercolors, except in the case of SC1604, for which we have ACS imaging. Since the ACS imaging is more precise, and the central wavelengths are comparable, we adopt those colors for that case. 

\begin{table*}
\caption{Spectroscopic and X-ray Catalog Information}
\label{srcsum}
\begin{tabular}{lrrrccc}
\footnotesize{LSS}
 & \footnotesize{Spectroscopic}
 & \footnotesize{Spectroscopic}
 & \footnotesize{X-ray Sources,}
 & \footnotesize{X-ray Sources,} & \footnotesize{Attempted}
 & \footnotesize{Confirmed}\\
\footnotesize{ }
 & \footnotesize{Targets}
 & \footnotesize{Redshifts$^a$}
 & \footnotesize{$>$3$\sigma$ ($>$2$\sigma$)$^b$}
 & \footnotesize{Matched$^c$}
 & \footnotesize{Redshifts$^d$}
 & \footnotesize{Redshifts$^a$}\\
\midrule
     RCS0224 &  \phantom{1}619 &  \phantom{1}507 & 119(138) &  \phantom{1}92(106) &   \phantom{1}7(8)\phantom{1} &   \phantom{1}7(8)\phantom{1} \\
      Cl0849 &  \phantom{1}926 &  \phantom{1}522 & 183(214) & 120(143) &  26(31) &  13(17) \\
     RXJ0910 &  \phantom{1}991 &  \phantom{1}750 & 218(248) & 125(141) &  43(46) &  27(30) \\
     RXJ1221 &  \phantom{1}681 &  \phantom{1}535 & 133(156) &  \phantom{1}81(96)\phantom{1} &  33(36) &  19(22) \\
      Cl1350 &  \phantom{1}828 &  \phantom{1}638 &  \phantom{1}85(101) &  \phantom{1}53(66)\phantom{1} &  23(27) &  17(21) \\
     RXJ1757 &  \phantom{1}970 &  \phantom{1}757 &  \phantom{1}84(98)\phantom{1} &  \phantom{1}59(65)\phantom{1} &  38(40) &  22(24) \\
      SC1604 & 2445 & 1849 & 144(179) &  101(125) &  44(50) &  27(30) \\
      SG0023 & 1155 &  \phantom{1}943 &  \phantom{1}92(109) &  \phantom{1}66(80)\phantom{1} &  43(51) &  29(35) \\
      SC1324 & 1690 & 1352 & 179(215) & 122(140) &  49(60) &  33(41) \\
     RXJ1821 &  \phantom{1}744 &  \phantom{1}626 &  \phantom{1}95(110) &  \phantom{1}63(72)\phantom{1} &  44(49) &  24(29) \\
     RXJ1716 &  \phantom{1}828 &  \phantom{1}571 &  \phantom{1}95(118) &  \phantom{1}80(94\phantom{1}) &  23(29) &  10(11) \\
     RXJ1053 &  \phantom{1}704 &  \phantom{1}410 & 115(131) &  \phantom{1}82(94)\phantom{1} &  22(25) &   \phantom{1}7(8)\phantom{1} \\
\bottomrule
\multicolumn{7}{l}{$^{\rm a}$ \footnotesize{Only includes high quality redshifts.}}\\
\multicolumn{7}{l}{$^{\rm b}$ \footnotesize{Includes sources with a significance $>$3$\sigma$ ($>$2$\sigma$) in at least one of the three bands: soft, hard, or full.}}\\
\multicolumn{7}{l}{$^{\rm c}$ \footnotesize{X-ray sources matched to optical counterparts.}}\\
\multicolumn{7}{p{14cm}}{$^{\rm d}$ \footnotesize{Includes all X-ray sources that were targeted for spectroscopy, regardless of the quality of measured redshift.}}
\end{tabular}
\end{table*}

Figure \ref{CMDS} shows CMDs for all LSSs in our sample. All spectroscopically confirmed supercluster/cluster members are shown. Diamonds indicate the confirmed X-ray AGNs within each LSS, which are analyzed in Section \ref{sec:AGN}. The red sequence for each LSS is delineated by dashed lines. Red sequence fits for each LSS were calculated using a linear fitting and $\sigma$-clipping technique. First, a fit to a linear model, of the form \begin{equation} C = C_0 + m \times B \end{equation}
where $C$ is either $M_{Blue} - M_{Red}$ or F606W-F814W and $B$  is either $M_{Red}$ or F814W, was carried out on member galaxies within a chosen magnitude and color range using a $\chi^2$ minimization \citep{gladders98,stott09}. The fit was initialized with a color range chosen ``by eye'' to conform to the apparent width of the red sequence of the LSS. The magnitude bounds were chosen to be where the red sequence was approximately spectroscopically complete for the entire sample, which we determined as $M_{Red}<-20.9$. In some cases, brightest cluster/group galaxies appearing at bluer colors had to be cut, as well. After the initial fit, the colors were normalized to remove the slope of the red sequence and the resultant color distribution was then fit to a single Gaussian using iterative $3\sigma$ clipping. At the conclusion of the algorithm, the boundaries of the red sequence were defined by a 3$\sigma\ $offset from the center, except for SC1604. For this supercluster, the color dispersion was inflated due to the large redshift extent of the LSS, and 2$\sigma$ offsets were used to achieve reasonable boundaries. 

While red sequences are well-defined for most of our LSSs, the size of the blue cloud varies considerably. We can quantify this by looking at the fraction of blue galaxies for each case. Here, we will define a blue galaxy as one below the lower boundary of the red sequence, including only galaxies with $M_{Red}<-20.9$. The blue fractions for each LSS are displayed in Table \ref{globspectab}. Unsurprisingly, two of the highest redshift LSSs, Cl0849 and RXJ1053, have the highest blue fractions. SG0023 and SC1604 have relatively high blue fractions as well, which are probably indicative of actively star-forming galaxy populations. SC1324 has a relatively high blue fraction as well. The other LSSs have lower blue fractions, which may indicate that their populations are more quiescent. 

RXJ1716 and RXJ0910 have blue fractions that may be lower than expected, especially considering the high redshift of RXJ0910. Their blue fractions are comparable to the most passive LSS populations, despite their spectral signatures indicating more actively star-forming galaxy populations (See Section \ref{sec:globspec}). The issue here may be insufficient completeness in the blue cloud. Indeed, when we include galaxies with only photometric redshifts in calculating the blue fractions, these LSSs have blue fractions of 27.8 \% and 34.6 \%, respectively. These can be compared to RXJ1757 and RXJ1821, which, as members of the Passive sub-sample, may be expected to have relatively low blue fractions based on their spectral properties. RXJ0910 in particular should have a higher blue fraction than these two LSSs considering its considerably higher redshift. We do find this to be the case, measuring blue fractions of 28.8 \% and 27.3 \% for RXJ1757 and RXJ1821, respectively, when galaxies with only photometric redshifts are included. This implies that RXJ1716 and RXJ0910 have more active star formation than the uncorrected blue fractions in Table \ref{globspectab} suggest.

\section{Spectral Properties}
\label{sec:globspec}

Eventually, we want to learn about the properties of the average AGN host galaxy in each LSS. First, we need a baseline to compare that to, which is the average properties of the average galaxy in the overall population. We can accomplish this using coadded spectra, through which we can examine the average star formation and starburst histories of our sample. We coadded all individual spectra from each LSS into a single composite spectrum, using the cutoff of $M_{Red}<-20.9$ described in Section \ref{TRSATBP}, according to the method of \citet{lemaux09,lemaux12}. We use three spectral features as our main probes of star formation history: the \OII\ doublet, at 3726-3729\AA, the H$\delta$ line, at 4102\AA, and the 4000\AA\ break. Each of these features probes a different star formation regime. The \OII\ emission line is often used as an indicator of current star formation, especially when the H$\alpha$ line is too redshifted for use, as with optical spectrography \citep{pogg99}. Caution must be exercised when using the \OII\ line, since \OII\ emission can also be created through LINER and Seyfert processes \citep[See, e.g.,][Lemaux et al. (2016, in prep.)]{yan06,lemaux10,koc12}. 

While populations of young O stars can create strong Balmer and \OII\ emission, galaxies with continua dominated instead by older A and B stars tend to exhibit H$\delta$ absorption \citep{pogg97}. Due to the lifetime of these stars, the H$\delta$ line is therefore an indicator of star formation in the past $\sim 1$ Gyr. The 4000\AA\ break, quantitatively measured as D$_{\rm n}$(4000), also measures time since the most recent star formation event, as it increases with the mean stellar age \citep{kauf03}, and is relatively insensitive to changes in dust and stellar-phase metallicity \citep{lemaux12}.

\begin{figure*}
\includegraphics[width=\textwidth]{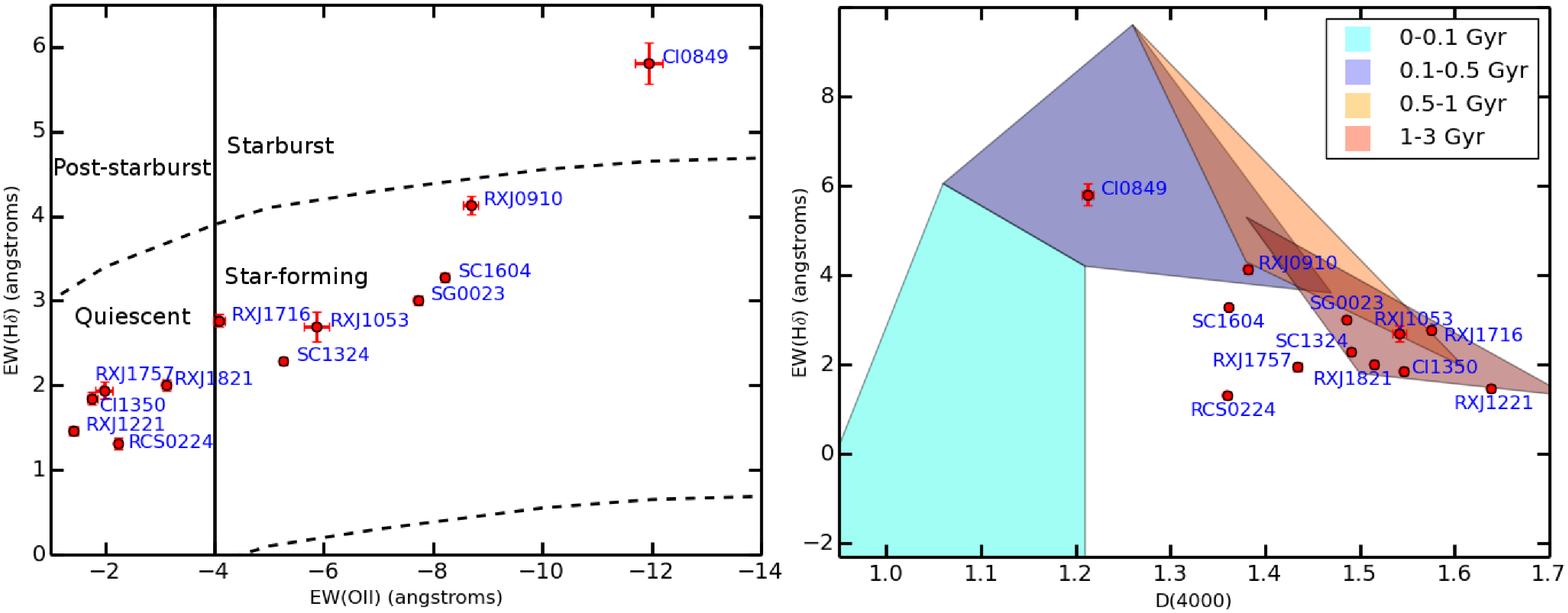}
\caption{
Measurements of spectral features for the composite spectra of each LSS in our sample are shown. {\it Left}: EW(\OII) and EW(\Hd) measurements are shown. The four regions on the plot are based on the spectral types of \citet{dress99} and \citet{pogg99}, using low redshift galaxies.  The dashed lines are derived from \citet{dress04}, and enclose 95\% of normal star-forming galaxies. {\it Right}: EW(\Hd) and \DFK\ measurements are shown. Additionally, regions are plotted which represent a range of post-starburst models derived from \citet{BC03} and \citet{bc07}. The models use either a Salpeter or Chabrier initial mass function, a stellar extinction of $0\le E_s\left(B-V\right)\le0.25$, and either stellar-phase metallicities of $Z=0.4Z_{\odot},Z_{\odot}$. Models have either a single burst, with exponential declining star formation ($\tau=10$ Myr) or include a secondary burst, occurring 2-3 Gyr after the initial starburst event, creating 5-100\% of the galaxy's stellar mass or new stars. 
}
\label{globspecfig}
\end{figure*}

Our measurements of D$_{\rm n}$(4000) and the \OII\ and H$\delta$ equivalent widths\footnote{Equivalent widths are measured using the bandpass method. Additionally, EW(H$\delta$) measurements are corrected for infill.} for the composite spectra of all LSSs are shown in Table \ref{globspectab}. Additionally, these are plotted in Figure \ref{globspecfig}. In the left panel, EW(\OII) is plotted versus EW(H$\delta$). The four regions on the plot, delineated by the vertical line and the upper dashed line, are based on the spectral types of \citet{dress99} and \citet{pogg99}, using low redshift galaxies. Starting from the lower right and moving clockwise, they correspond to star-forming galaxies, quiescent galaxies, post-starburst galaxies, and starbursting galaxies. The dashed lines are derived from \citet{dress04}, and enclose 95\% of normal star-forming galaxies. In the right panel, D$_{\rm n}$(4000) is plotted versus EW(H$\delta$). The shaded regions are areas in this phase space spanned by a range of models derived from \citet{BC03} and \citet{bc07}. The models use either a Salpeter or Chabrier initial mass function, a stellar extinction of $0\le Es\left(B-V\right)\le0.25$, and have stellar-phase metallicites of $Z=0.4Z_{\odot},Z_{\odot}$. The models have either a single burst, or a secondary burst of varing intensity. The four different shaded regions represent models with differing times since the last starburst event, shown in the legend. 

With these spectral measurements, the LSSs in our sample are differentiated into several sub-samples. The composites of the highest redshift LSSs have the highest average EW(\OII) and EW(H$\delta$). On average, these LSSs appear to have the highest levels of star formation in our sample, and the average member galaxy is likely currently starbursting. This is supported by the D$_{\rm n}$(4000) measurements, where the average member galaxies of these LSSs have some of the lowest times since the last starburst event. This sub-sample is followed by SG0023 and SC1604, which also appear to have a galaxy population which is star-forming, on average, although at lower levels than the high-redshift sub-sample, and these populations have shorter average times since the last starburst event. 

RCS0224, RXJ1221, Cl1350, RXJ1757, and RXJ1821 make up a sub-sample of LSSs with quiescent galaxy populations. With the lowest EW(\OII) and EW(H$\delta$) and the largest D$_{\rm n}$(4000) values, the galaxy populations of these LSSs seem to be passive on average and have the longest times since the last starburst event. In an intermediate range are SC1324 and RXJ1716. The average EW(H$\delta$) measurements for members of these LSSs lie near the boundary between quiescent and star-forming. Additionally, the average time since last starburst appears to be in between those for the actively more actively star-forming SG0023 and SC1604 LSSs and the passive sub-sample of LSSs. 

\begin{table*}
\caption{LSS Average Spectral Properties and Blue Fractions}
\label{globspectab}
\begin{tabular}{lcccc}
\footnotesize{LSS}
 & \footnotesize{EW(\OII)}
 & \footnotesize{EW(\Hd)}
 & \footnotesize{\DFK}
 & \footnotesize{Blue Fraction$^a$}\\
\footnotesize{}
 & \footnotesize{\AA}
 & \footnotesize{\AA}
 & \footnotesize{}
 & \footnotesize{}\\
\midrule
SG0023 & \phantom{1}-7.7 $\pm$ 0.1 & 3.0 $\pm$ 0.1 & 1.486 $\pm$ 0.002 & 0.35\\
RCS0224 & \phantom{1}-2.2 $\pm$ 0.1 & 1.3 $\pm$ 0.1 & 1.360 $\pm$ 0.002 & 0.27\\
Cl0849 & -11.9 $\pm$ 0.3 & 5.8 $\pm$ 0.2 & 1.213 $\pm$ 0.006 & 0.51\\
RXJ0910 & \phantom{1}-8.7$\pm$ 0.1 & 4.1 $\pm$ 0.1 & 1.382 $\pm$ 0.004 & 0.26\\
RXJ1053 & \phantom{1}-5.9$\pm$ 0.2 & 2.7 $\pm$ 0.2 & 1.541 $\pm$ 0.007 & 0.42\\
RXJ1221 & \phantom{1}-1.4$\pm$ 0.1 & 1.5 $\pm$ 0.1 & 1.638 $\pm$ 0.003 & 0.19\\
SC1324 & \phantom{1}-5.3 $\pm$ 0.1 & 2.28 $\pm$ 0.04 & 1.491 $\pm$ 0.002 & 0.32\\
Cl1350 & \phantom{1}-1.8$\pm$ 0.1 & 1.8 $\pm$ 0.1 & 1.546 $\pm$ 0.003 & 0.23\\
SC1604 & \phantom{1}-8.2$\pm$ 0.1 & 3.3 $\pm$ 0.1 & 1.352 $\pm$ 0.002 & 0.38\\
RXJ1716 & \phantom{1}-4.1$\pm$ 0.1 & 2.8 $\pm$ 0.1 & 1.575 $\pm$ 0.003 & 0.18\\
RXJ1757 & \phantom{1}-2.0$\pm$ 0.2 & 1.9 $\pm$ 0.1 & 1.434 $\pm$ 0.004 & 0.16\\
RXJ1821 & \phantom{1}-3.1$\pm$ 0.1 & 2.0 $\pm$ 0.1 & 1.515 $\pm$ 0.003 & 0.23\\
\bottomrule
\multicolumn{5}{p{10cm}}{$^{\rm a}$ \footnotesize{Blue fraction is defined here as the fraction of galaxies below the bounds of the red sequence, and with $M_{Red}<-20.9$. See Section \ref{TRSATBP} for more details.}}
\end{tabular}
\end{table*}

\section{Summary of Global Characteristics}
\label{sec:SGC}

Our analysis of the optical colors and average spectral properties suggests that our sample can be roughly broken up into four sub-samples of LSSs that are roughly similar. The highest redshift LSSs in our sample (Cl0849, RXJ0910, RXJ1053) are the bluest and have the most actively star-forming galaxy populations on average. In fact, significant fractions of actively starbursting galaxies may be present. Hereafter, we will refer to this sub-sample as the High-$z$ LSSs. The galaxy populations of SG0023 \& SC1604, similarly, are relatively blue and actively forming stars, on average, although not quite to the extent of the High-$z$ LSSs. While there may appear to be similarities between these sub-samples, our later X-ray analysis shows marked differences between the AGNs in SG0023 \& SC1604 and the High-$z$ LSSs, justifying a separate analysis. 

Five of the LSSs (RCS0224, RXJ1221, Cl1350, RXJ1757, and RXJ1821) appear to house the most quiescent member populations based on our spectral measurements, and they have the lowest blue fractions. We will call this sub-sample the Passive LSSs\footnote{Note that `Passive' is used here in a relative sense. The galaxy populations of these LSSs have less star formation activity than the rest of our sample, on average. We do not mean to imply that the average galaxy in these LSSs has negligible star formation.}. This leaves SC1324 and RXJ1716. As mentioned earlier, the spectral properties of these LSSs place their star formation activity at a level higher than the average galaxy in the Passive LSSs, but still lower than SG0023 \& SC1604. The higher blue fraction in SC1324 also seems to show it may have more active star formation than the Passive sub-sample. While RXJ1716 seems to have a rather low blue fraction, correcting for completeness using photometric redshifts implies it has a blue fraction at least no lower than the Passive sub-sample (see Section \ref{TRSATBP}). We analyze SC1324 and RXJ1716 together, and refer to them as the Intermediate sub-sample. 

\section{AGN and Host Properties}
\label{sec:AGN}

Using optical sources with redshifts with quality flags of $Q=3$ or 4
and the results of our optical matching, we were able to identify
X-ray sources that are members of the LSSs in our
sample (see Table \ref{srcsum}). In total, we find 61 AGNs across all LSSs. 18 of these are in the Passive sub-sample, 12 are in the Intermediate sub-sample, 7 in SG0023, 10 in SC1604, and 14 are in the High-$z$ sub-sample. 
Note that
these numbers include all sources detected at a $>$2$\sigma$
level in at least one of the three X-ray passbands (see Table
\ref{AGNtab}), and seven out of the 61 AGNs were detected at lower than the 3$\sigma$ level. 

\subsection{Spatial Distribution}
\label{sec:spatdist}

\begin{table*}
\caption{AGN Summary}
\label{AGNtab}
\bgroup
\def\arraystretch{0.98}
\begin{tabular}{llllllllllr}
\scriptsize{LSS}
 & \scriptsize{Num.}
 & \scriptsize{R.A. (J2000)}
 & \scriptsize{Dec. (J2000)} 
 & \scriptsize{$z$}
 & \scriptsize{$L_x$ (soft)$^a$}
 & \scriptsize{$L_x$ (hard)$^a$}
 & \scriptsize{$L_x$ (full)$^a$}
 & \scriptsize{Detection Sig.($\sigma$)}
 & \scriptsize{$p^b$}
 & \scriptsize{RSO$^c$}
 \\
\midrule
 \footnotesize{SG0023} & \footnotesize{ 1} & \footnotesize{00\ 24\ 10.9} & \footnotesize{+04\ 29\ 23.5} & \footnotesize{0.823} & \footnotesize{6.2} & \footnotesize{51.0} & \footnotesize{43.4} & \footnotesize{12.28} & \footnotesize{29.67} & \footnotesize{1.89} \\
\footnotesize{SG0023} & \footnotesize{ 2} & \footnotesize{00\ 24\ 15.5} & \footnotesize{+04\ 23\ 09.1} & \footnotesize{0.829} & \footnotesize{8.3} & \footnotesize{$^d$} & \footnotesize{19.8} & \footnotesize{9.64} & \footnotesize{7.11} & \footnotesize{-0.71} \\
\footnotesize{SG0023} & \footnotesize{ 3} & \footnotesize{00\ 23\ 54.9} & \footnotesize{+04\ 25\ 24.6} & \footnotesize{0.830} & \footnotesize{0.7} & \footnotesize{7.1} & \footnotesize{5.5} & \footnotesize{3.37} & \footnotesize{10.41} & \footnotesize{-0.28} \\
\footnotesize{SG0023} & \footnotesize{ 4} & \footnotesize{00\ 24\ 09.4} & \footnotesize{+04\ 22\ 41.1} & \footnotesize{0.841} & \footnotesize{25.0} & \footnotesize{36.2} & \footnotesize{77.1} & \footnotesize{33.20} & \footnotesize{0.36} & \footnotesize{6.28} \\
\footnotesize{SG0023} & \footnotesize{ 5} & \footnotesize{00\ 23\ 52.2} & \footnotesize{+04\ 22\ 59.8} & \footnotesize{0.844} & \footnotesize{25.7} & \footnotesize{109.8} & \footnotesize{119.5} & \footnotesize{50.01} & \footnotesize{0.20} & \footnotesize{3.09} \\
\footnotesize{SG0023} & \footnotesize{ 6} & \footnotesize{00\ 23\ 45.6} & \footnotesize{+04\ 22\ 59.4} & \footnotesize{0.850} & \footnotesize{9.7} & \footnotesize{21.4} & \footnotesize{37.5} & \footnotesize{19.84} & \footnotesize{0.68} & \footnotesize{4.91} \\
\footnotesize{SG0023} & \footnotesize{ 7} & \footnotesize{00\ 24\ 07.6} & \footnotesize{+04\ 27\ 26.9} & \footnotesize{0.854} & \footnotesize{3.3} & \footnotesize{$^d$} & \footnotesize{10.5} & \footnotesize{4.16} & \footnotesize{12.36} & \footnotesize{2.00} \\
\footnotesize{RCS0224} & \footnotesize{ 1} & \footnotesize{02\ 24\ 36.9} & \footnotesize{-01\ 53\ 56.5} & \footnotesize{0.774} & \footnotesize{7.8} & \footnotesize{25.4} & \footnotesize{30.6} & \footnotesize{45.19} & \footnotesize{0.19} & \footnotesize{7.68} \\
\footnotesize{RCS0224} & \footnotesize{ 2} & \footnotesize{02\ 24\ 38.2} & \footnotesize{-01\ 53\ 49.7} & \footnotesize{0.778} & \footnotesize{0.8} & \footnotesize{$^d$} & \footnotesize{2.1} & \footnotesize{3.13} & \footnotesize{0.01} & \footnotesize{-0.11} \\
\footnotesize{RCS0224} & \footnotesize{ 3} & \footnotesize{02\ 24\ 37.3} & \footnotesize{-01\ 55\ 02.4} & \footnotesize{0.778} & \footnotesize{0.2} & \footnotesize{29.2} & \footnotesize{11.5} & \footnotesize{24.68} & \footnotesize{0.02} & \footnotesize{1.66} \\
\footnotesize{RCS0224} & \footnotesize{ 4} & \footnotesize{02\ 24\ 39.8} & \footnotesize{-01\ 54\ 30.7} & \footnotesize{0.781} & \footnotesize{0.8} & \footnotesize{2.4} & \footnotesize{3.5} & \footnotesize{5.20} & \footnotesize{0.20} & \footnotesize{-0.22} \\
\footnotesize{Cl0849} & \footnotesize{ 1} & \footnotesize{08\ 49\ 04.1} & \footnotesize{+44\ 56\ 47.3} & \footnotesize{1.263} & \footnotesize{1.3} & \footnotesize{16.0} & \footnotesize{10.5} & \footnotesize{9.79} & \footnotesize{0.62} & \footnotesize{0.49} \\
\footnotesize{Cl0849} & \footnotesize{ 2} & \footnotesize{08\ 48\ 58.8} & \footnotesize{+44\ 56\ 21.8} & \footnotesize{1.263} & \footnotesize{1.8} & \footnotesize{9.8} & \footnotesize{10.5} & \footnotesize{10.21} & \footnotesize{0.58} & \footnotesize{0.53} \\
\footnotesize{Cl0849} & \footnotesize{ 3} & \footnotesize{08\ 49\ 05.3} & \footnotesize{+44\ 52\ 04.4} & \footnotesize{1.265} & \footnotesize{2.1} & \footnotesize{56.7} & \footnotesize{31.9} & \footnotesize{24.50} & \footnotesize{2.28} & \footnotesize{1.74} \\
\footnotesize{Cl0849} & \footnotesize{ 4} & \footnotesize{08\ 49\ 03.9} & \footnotesize{+44\ 50\ 25.2} & \footnotesize{1.274} & \footnotesize{1.2} & \footnotesize{11.3} & \footnotesize{8.7} & \footnotesize{6.19} & \footnotesize{4.87} & \footnotesize{-0.48} \\
\footnotesize{RXJ0910} & \footnotesize{ 1} & \footnotesize{09\ 10\ 40.9} & \footnotesize{+54\ 20\ 08.8} & \footnotesize{1.096} & \footnotesize{$^d$} & \footnotesize{10.9} & \footnotesize{4.9} & \footnotesize{6.06} & \footnotesize{0.99} & \footnotesize{2.15} \\
\footnotesize{RXJ0910} & \footnotesize{ 2} & \footnotesize{09\ 10\ 48.2} & \footnotesize{+54\ 22\ 30.0} & \footnotesize{1.097} & \footnotesize{1.0} & \footnotesize{$^d$} & \footnotesize{3.8} & \footnotesize{4.35} & \footnotesize{0.19} & \footnotesize{-0.25} \\
\footnotesize{RXJ0910} & \footnotesize{ 3} & \footnotesize{09\ 10\ 42.8} & \footnotesize{+54\ 20\ 11.0} & \footnotesize{1.099} & \footnotesize{0.7} & \footnotesize{7.4} & \footnotesize{5.1} & \footnotesize{6.57} & \footnotesize{0.52} & \footnotesize{0.24} \\
\footnotesize{RXJ0910} & \footnotesize{ 4} & \footnotesize{09\ 09\ 54.0} & \footnotesize{+54\ 17\ 55.1} & \footnotesize{1.101} & \footnotesize{$^d$} & \footnotesize{22.7} & \footnotesize{13.3} & \footnotesize{8.10} & \footnotesize{0.45} & \footnotesize{1.54} \\
\footnotesize{RXJ0910} & \footnotesize{ 5} & \footnotesize{09\ 10\ 34.9} & \footnotesize{+54\ 24\ 54.9} & \footnotesize{1.102} & \footnotesize{1.0} & \footnotesize{$^d$} & \footnotesize{2.7} & \footnotesize{2.79} & \footnotesize{0.14} & \footnotesize{0.01} \\
\footnotesize{RXJ0910} & \footnotesize{ 6} & \footnotesize{09\ 09\ 55.7} & \footnotesize{+54\ 18\ 14.6} & \footnotesize{1.102} & \footnotesize{2.5} & \footnotesize{20.7} & \footnotesize{16.5} & \footnotesize{10.68} & \footnotesize{0.01} & \footnotesize{2.02} \\
\footnotesize{RXJ0910} & \footnotesize{ 7} & \footnotesize{09\ 09\ 45.2} & \footnotesize{+54\ 16\ 33.9} & \footnotesize{1.105} & \footnotesize{$^d$} & \footnotesize{13.4} & \footnotesize{7.7} & \footnotesize{3.34} & \footnotesize{1.84} & \footnotesize{1.13} \\
\footnotesize{RXJ0910} & \footnotesize{ 8} & \footnotesize{09\ 10\ 42.8} & \footnotesize{+54\ 20\ 37.2} & \footnotesize{1.106} & \footnotesize{0.7} & \footnotesize{8.2} & \footnotesize{6.7} & \footnotesize{8.38} & \footnotesize{0.65} & \footnotesize{2.05} \\
\footnotesize{RXJ0910} & \footnotesize{ 9} & \footnotesize{09\ 10\ 04.9} & \footnotesize{+54\ 17\ 36.6} & \footnotesize{1.117} & \footnotesize{1.0} & \footnotesize{$^d$} & \footnotesize{5.2} & \footnotesize{3.52} & \footnotesize{3.77} & \footnotesize{0.44} \\
\footnotesize{RXJ1053} & \footnotesize{ 1} & \footnotesize{10\ 53\ 39.6} & \footnotesize{+57\ 36\ 48.6} & \footnotesize{1.125} & \footnotesize{1.8} & \footnotesize{$^d$} & \footnotesize{5.5} & \footnotesize{4.99} & \footnotesize{5.23} & \footnotesize{5.77} \\
\footnotesize{RXJ1221} & \footnotesize{ 1} & \footnotesize{12\ 21\ 51.4} & \footnotesize{+49\ 19\ 29.7} & \footnotesize{0.692} & \footnotesize{1.2} & \footnotesize{9.1} & \footnotesize{7.6} & \footnotesize{8.16} & \footnotesize{3.55} & \footnotesize{1.72} \\
\footnotesize{RXJ1221} & \footnotesize{ 2} & \footnotesize{12\ 21\ 49.0} & \footnotesize{+49\ 12\ 49.0} & \footnotesize{0.696} & \footnotesize{0.4} & \footnotesize{2.9} & \footnotesize{2.4} & \footnotesize{3.33} & \footnotesize{1.37} & \footnotesize{3.88} \\
\footnotesize{RXJ1221} & \footnotesize{ 3} & \footnotesize{12\ 21\ 00.9} & \footnotesize{+49\ 20\ 38.7} & \footnotesize{0.698} & \footnotesize{$^d$} & \footnotesize{18.1} & \footnotesize{8.9} & \footnotesize{7.98} & \footnotesize{1.23} & \footnotesize{1.76} \\
\footnotesize{RXJ1221} & \footnotesize{ 4} & \footnotesize{12\ 21\ 56.1} & \footnotesize{+49\ 13\ 45.7} & \footnotesize{0.699} & \footnotesize{$^d$} & \footnotesize{5.9} & \footnotesize{3.0} & \footnotesize{3.67} & \footnotesize{0.25} & \footnotesize{0.97} \\
\footnotesize{SC1324} & \footnotesize{ 1} & \footnotesize{13\ 24\ 51.5} & \footnotesize{+30\ 12\ 42.6} & \footnotesize{0.660} & \footnotesize{1.1} & \footnotesize{$^d$} & \footnotesize{2.2} & \footnotesize{2.51} & \footnotesize{7.48} & \footnotesize{2.12} \\
\footnotesize{SC1324} & \footnotesize{ 2} & \footnotesize{13\ 24\ 36.6} & \footnotesize{+30\ 23\ 17.4} & \footnotesize{0.662} & \footnotesize{1.1} & \footnotesize{8.0} & \footnotesize{6.8} & \footnotesize{4.28} & \footnotesize{27.25} & \footnotesize{2.02} \\
\footnotesize{SC1324} & \footnotesize{ 3} & \footnotesize{13\ 24\ 38.6} & \footnotesize{+30\ 58\ 08.3} & \footnotesize{0.676} & \footnotesize{2.8} & \footnotesize{$^d$} & \footnotesize{7.6} & \footnotesize{5.20} & \footnotesize{5.49} & \footnotesize{2.15} \\
\footnotesize{SC1324} & \footnotesize{ 4} & \footnotesize{13\ 24\ 52.9} & \footnotesize{+30\ 52\ 18.0} & \footnotesize{0.697} & \footnotesize{1.3} & \footnotesize{$^d$} & \footnotesize{3.0} & \footnotesize{2.64} & \footnotesize{0.97} & \footnotesize{1.67} \\
\footnotesize{SC1324} & \footnotesize{ 5} & \footnotesize{13\ 24\ 52.0} & \footnotesize{+30\ 50\ 51.5} & \footnotesize{0.700} & \footnotesize{$^d$} & \footnotesize{13.2} & \footnotesize{6.6} & \footnotesize{6.47} & \footnotesize{4.30} & \footnotesize{1.09} \\
\footnotesize{SC1324} & \footnotesize{ 6} & \footnotesize{13\ 24\ 50.6} & \footnotesize{+30\ 56\ 24.2} & \footnotesize{0.702} & \footnotesize{0.7} & \footnotesize{5.7} & \footnotesize{4.2} & \footnotesize{3.10} & \footnotesize{1.49} & \footnotesize{0.36} \\
\footnotesize{SC1324} & \footnotesize{ 7} & \footnotesize{13\ 25\ 08.7} & \footnotesize{+30\ 52\ 13.5} & \footnotesize{0.757} & \footnotesize{1.5} & \footnotesize{$^d$} & \footnotesize{4.4} & \footnotesize{2.69} & \footnotesize{2.15} & \footnotesize{0.94} \\
\footnotesize{SC1324} & \footnotesize{ 8} & \footnotesize{13\ 24\ 28.8} & \footnotesize{+30\ 53\ 19.5} & \footnotesize{0.778} & \footnotesize{1.5} & \footnotesize{$^d$} & \footnotesize{4.4} & \footnotesize{2.62} & \footnotesize{51.71} & \footnotesize{0.75} \\
\footnotesize{Cl1350} & \footnotesize{ 1} & \footnotesize{13\ 49\ 34.4} & \footnotesize{+60\ 02\ 28.1} & \footnotesize{0.798} & \footnotesize{2.6} & \footnotesize{65.6} & \footnotesize{38.5} & \footnotesize{12.95} & \footnotesize{1.09} & \footnotesize{1.11} \\
\footnotesize{Cl1350} & \footnotesize{ 2} & \footnotesize{13\ 50\ 46.0} & \footnotesize{+60\ 07\ 00.9} & \footnotesize{0.802} & \footnotesize{$^d$} & \footnotesize{18.0} & \footnotesize{10.5} & \footnotesize{7.01} & \footnotesize{0.04} & \footnotesize{0.71} \\
\footnotesize{Cl1350} & \footnotesize{ 3} & \footnotesize{13\ 50\ 50.1} & \footnotesize{+60\ 08\ 03.3} & \footnotesize{0.807} & \footnotesize{2.2} & \footnotesize{3.8} & \footnotesize{7.4} & \footnotesize{5.21} & \footnotesize{0.34} & \footnotesize{-0.69} \\
\footnotesize{SC1604} & \footnotesize{ 1} & \footnotesize{16\ 04\ 23.9} & \footnotesize{+43\ 11\ 25.8} & \footnotesize{0.867} & \footnotesize{12.7} & \footnotesize{24.1} & \footnotesize{47.2} & \footnotesize{28.11} & \footnotesize{0.63} & \footnotesize{4.30} \\
\footnotesize{SC1604} & \footnotesize{ 2} & \footnotesize{16\ 04\ 25.9} & \footnotesize{+43\ 12\ 45.3} & \footnotesize{0.871} & \footnotesize{6.3} & \footnotesize{11.3} & \footnotesize{21.1} & \footnotesize{11.78} & \footnotesize{1.02} & \footnotesize{1.42} \\
\footnotesize{SC1604} & \footnotesize{ 3} & \footnotesize{16\ 04\ 24.0} & \footnotesize{+43\ 04\ 35.1} & \footnotesize{0.899} & \footnotesize{$^d$} & \footnotesize{$^d$} & \footnotesize{5.9} & \footnotesize{2.21} & \footnotesize{0.07} & \footnotesize{-0.11} \\
\footnotesize{SC1604} & \footnotesize{ 4} & \footnotesize{16\ 04\ 15.6} & \footnotesize{+43\ 10\ 16.6} & \footnotesize{0.900} & \footnotesize{17.2} & \footnotesize{37.5} & \footnotesize{62.7} & \footnotesize{13.16} & \footnotesize{1.55} & \footnotesize{1.76} \\
\footnotesize{SC1604} & \footnotesize{ 5} & \footnotesize{16\ 04\ 37.7} & \footnotesize{+43\ 08\ 58.0} & \footnotesize{0.900} & \footnotesize{1.1} & \footnotesize{6.5} & \footnotesize{6.2} & \footnotesize{3.02} & \footnotesize{1.59} & \footnotesize{$^e$}\\
\footnotesize{SC1604} & \footnotesize{ 6} & \footnotesize{16\ 04\ 06.1} & \footnotesize{+43\ 18\ 07.7} & \footnotesize{0.913} & \footnotesize{13.4} & \footnotesize{27.7} & \footnotesize{45.8} & \footnotesize{22.64} & \footnotesize{2.31} & \footnotesize{0.99} \\
\footnotesize{SC1604} & \footnotesize{ 7} & \footnotesize{16\ 04\ 36.7} & \footnotesize{+43\ 21\ 41.1} & \footnotesize{0.923} & \footnotesize{6.3} & \footnotesize{19.4} & \footnotesize{26.5} & \footnotesize{10.50} & \footnotesize{0.04} & \footnotesize{3.52} \\
\footnotesize{SC1604} & \footnotesize{ 8} & \footnotesize{16\ 04\ 01.4} & \footnotesize{+43\ 13\ 51.1} & \footnotesize{0.927} & \footnotesize{11.2} & \footnotesize{31.8} & \footnotesize{44.8} & \footnotesize{19.32} & \footnotesize{0.97} & \footnotesize{$^e$} \\
\footnotesize{SC1604} & \footnotesize{ 9} & \footnotesize{16\ 04\ 05.2} & \footnotesize{+43\ 15\ 20.8} & \footnotesize{0.934} & \footnotesize{3.2} & \footnotesize{$^d$} & \footnotesize{10.3} & \footnotesize{5.12} & \footnotesize{0.00} & \footnotesize{1.01} \\
\footnotesize{SC1604} & \footnotesize{10} & \footnotesize{16\ 04\ 10.9} & \footnotesize{+43\ 21\ 11.2} & \footnotesize{0.935} & \footnotesize{1.1} & \footnotesize{11.9} & \footnotesize{8.9} & \footnotesize{3.50} & \footnotesize{0.13} & \footnotesize{-0.97} \\
\footnotesize{RXJ1716} & \footnotesize{ 1} & \footnotesize{17\ 16\ 37.8} & \footnotesize{+67\ 07\ 30.3} & \footnotesize{0.805} & \footnotesize{2.7} & \footnotesize{5.3} & \footnotesize{9.2} & \footnotesize{5.26} & \footnotesize{0.40} & \footnotesize{0.09} \\
\footnotesize{RXJ1716} & \footnotesize{ 2} & \footnotesize{17\ 17\ 03.9} & \footnotesize{+67\ 08\ 59.2} & \footnotesize{0.809} & \footnotesize{$^d$} & \footnotesize{7.9} & \footnotesize{5.1} & \footnotesize{3.68} & \footnotesize{0.14} & \footnotesize{2.14} \\
\footnotesize{RXJ1716} & \footnotesize{ 3} & \footnotesize{17\ 16\ 24.4} & \footnotesize{+67\ 05\ 28.6} & \footnotesize{0.816} & \footnotesize{$^d$} & \footnotesize{9.9} & \footnotesize{4.9} & \footnotesize{2.44} & \footnotesize{0.74} & \footnotesize{1.95} \\
\footnotesize{RXJ1716} & \footnotesize{ 4} & \footnotesize{17\ 16\ 41.9} & \footnotesize{+67\ 06\ 07.0} & \footnotesize{0.821} & \footnotesize{$^d$} & \footnotesize{8.2} & \footnotesize{5.7} & \footnotesize{3.06} & \footnotesize{0.97} & \footnotesize{0.09} \\
\footnotesize{RXJ1757} & \footnotesize{ 1} & \footnotesize{17\ 57\ 25.2} & \footnotesize{+66\ 31\ 50.7} & \footnotesize{0.693} & \footnotesize{2.2} & \footnotesize{7.1} & \footnotesize{8.7} & \footnotesize{7.17} & \footnotesize{0.03} & \footnotesize{2.57} \\
\footnotesize{RXJ1757} & \footnotesize{ 2} & \footnotesize{17\ 57\ 07.5} & \footnotesize{+66\ 30\ 18.6} & \footnotesize{0.707} & \footnotesize{2.5} & \footnotesize{4.1} & \footnotesize{8.4} & \footnotesize{6.97} & \footnotesize{4.68} & \footnotesize{5.89} \\
\footnotesize{RXJ1821} & \footnotesize{ 1} & \footnotesize{18\ 21\ 07.7} & \footnotesize{+68\ 23\ 39.3} & \footnotesize{0.813} & \footnotesize{2.1} & \footnotesize{$^d$} & \footnotesize{6.2} & \footnotesize{2.76} & \footnotesize{0.73} & \footnotesize{0.04} \\
\footnotesize{RXJ1821} & \footnotesize{ 2} & \footnotesize{18\ 21\ 36.6} & \footnotesize{+68\ 30\ 02.1} & \footnotesize{0.814} & \footnotesize{3.2} & \footnotesize{$^d$} & \footnotesize{9.1} & \footnotesize{5.46} & \footnotesize{0.27} & \footnotesize{2.65} \\
\footnotesize{RXJ1821} & \footnotesize{ 3} & \footnotesize{18\ 21\ 02.8} & \footnotesize{+68\ 25\ 07.3} & \footnotesize{0.820} & \footnotesize{1.2} & \footnotesize{10.7} & \footnotesize{8.7} & \footnotesize{4.62} & \footnotesize{0.48} & \footnotesize{2.47} \\
\footnotesize{RXJ1821} & \footnotesize{ 4} & \footnotesize{18\ 21\ 23.9} & \footnotesize{+68\ 26\ 32.9} & \footnotesize{0.822} & \footnotesize{2.0} & \footnotesize{$^d$} & \footnotesize{7.7} & \footnotesize{4.42} & \footnotesize{0.37} & \footnotesize{0.59} \\
\footnotesize{RXJ1821} & \footnotesize{ 5} & \footnotesize{18\ 21\ 27.0} & \footnotesize{+68\ 32\ 34.7} & \footnotesize{0.824} & \footnotesize{8.2} & \footnotesize{14.1} & \footnotesize{25.9} & \footnotesize{12.15} & \footnotesize{1.40} & \footnotesize{0.22} \\
\bottomrule
\multicolumn{11}{p{17cm}}{$^{\rm a}$\scriptsize{Rest frame X-ray luminosity in units of 10$^{42}$ ergs s$^{-1}$. Soft, hard, and full bands are defined as 0.5-2.0, 2.0-7.0, and 0.5-7.0 keV, respectively.}}\\
\multicolumn{11}{p{17cm}}{$^{\rm b}$\scriptsize{Phase space metric, defined in Section \ref{sec:spatdist} as the product of the scaled distances on the sky and in velocity space to nearest group or cluster.}}\\
\multicolumn{11}{p{17cm}}{$^{\rm c}$\scriptsize{Scaled offset from the center of the red sequence fit in units of red sequence half-widths. Refer to Sections \ref{TRSATBP} and \ref{sec:hgalcolan}.}}\\
\multicolumn{11}{l}{$^{\rm d}$\scriptsize{Undetected in respective band.}}\\
\multicolumn{11}{p{17cm}}{$^{\rm e}$\scriptsize{Red sequence offsets for SC1604 AGN hosts were only calculated for those with ACS imaging.}}
\end{tabular}
\egroup
\end{table*}

Examining the spatial distribution of AGNs located within each cluster
can give insight into what processes triggered their nuclear
activity. Their positions and characteristics are given
in Table \ref{AGNtab}.

We find AGNs distributed across our LSSs. They are represented in dense cluster cores, in galaxy groups, on the outskirts of groups and clusters, and within the redshift bounds of our LSSs, but not associated with any particular cluster or group. These different regimes can be delineated and analyzed using a phase space metric that accounts for separations both on the sky and in radial velocity. 

\begin{figure}
\includegraphics[width=0.5\textwidth]{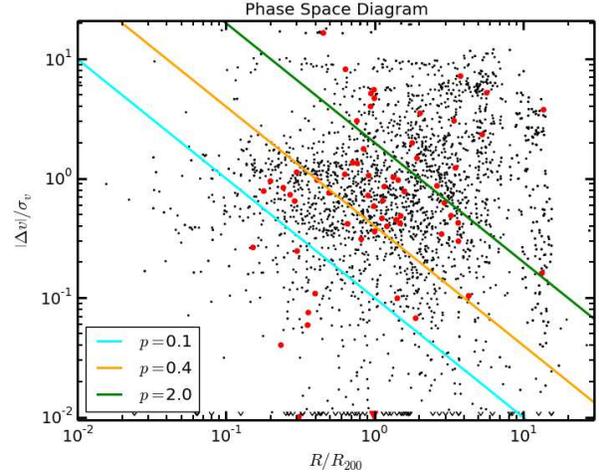}
\caption{
For each point, distance to the nearest cluster is plotted on the sky versus the difference in radial velocity. The nearest cluster is defined as the one with the lowest value of $r/r_{200}\times \left|\Delta v\right|/\sigma_v$. See Section \ref{sec:spatdist} for more details. AGNs are plotted with small circles, while all spectroscopically confirmed members of LSSs in our sample are plotted with small circles. AGNs and all sources that lie below the field of view of the plot are shown at the bottom with triangles and carets, respectively. Three lines of constant $p=r/r_{200}\times \left|\Delta v\right|/\sigma_v$ are plotted. 
}
\label{PSDfig}
\end{figure}

In Figure \ref{PSDfig}, we plot $r_{norm}=r/r_{200}$ versus $v_{norm}=\left|\Delta v\right|/\sigma_v$ for all spectroscopically confirmed members of our LSSs. Here, $r$ is the the projected distance to a group or cluster, while $r_{200}$ is for that same group or cluster. $\Delta v$ is the difference in line-of-sight velocity between that galaxy and a group or cluster, while $\sigma_v$ is the velocity dispersion for that same group or cluster, calculated within 0.5 Mpc of its center (See \citet{ascaso14} for details on calculating velocity dispersions.). It has been shown that, on diagrams such as these, lines of constant $p=r_{norm}\times v_{norm}$, sometimes called caustic lines, are useful for separating out different cluster populations, such as members in the virialized cores and infalling populations \citep[See, e.g., ][]{mamon04,gill05,haines12,noble13}. \citet{haines12} stacked 30 clusters from the Millenium Simulation, finding that galaxies that had been accreted into the clusters at earlier times, which should include mostly galaxies in the cluster cores, tended to be located inside caustic lines with lower values of $p$, while those accreted at later times, which should include infalling populations, were located outside caustic lines, having higher values of $p$. 

For each galaxy, $p$ must be calculated relative to a specific group or cluster. Since our sample contains a substantial subset of galaxies that are not close to any group or cluster, it is not always clear which object should be used for these references. To determine which group or cluster should be used for reference, $p$ was calculated for each galaxy relative to all groups and clusters, and the smallest value for $p$ was retained. 

\begin{figure}
\includegraphics[width=0.5\textwidth]{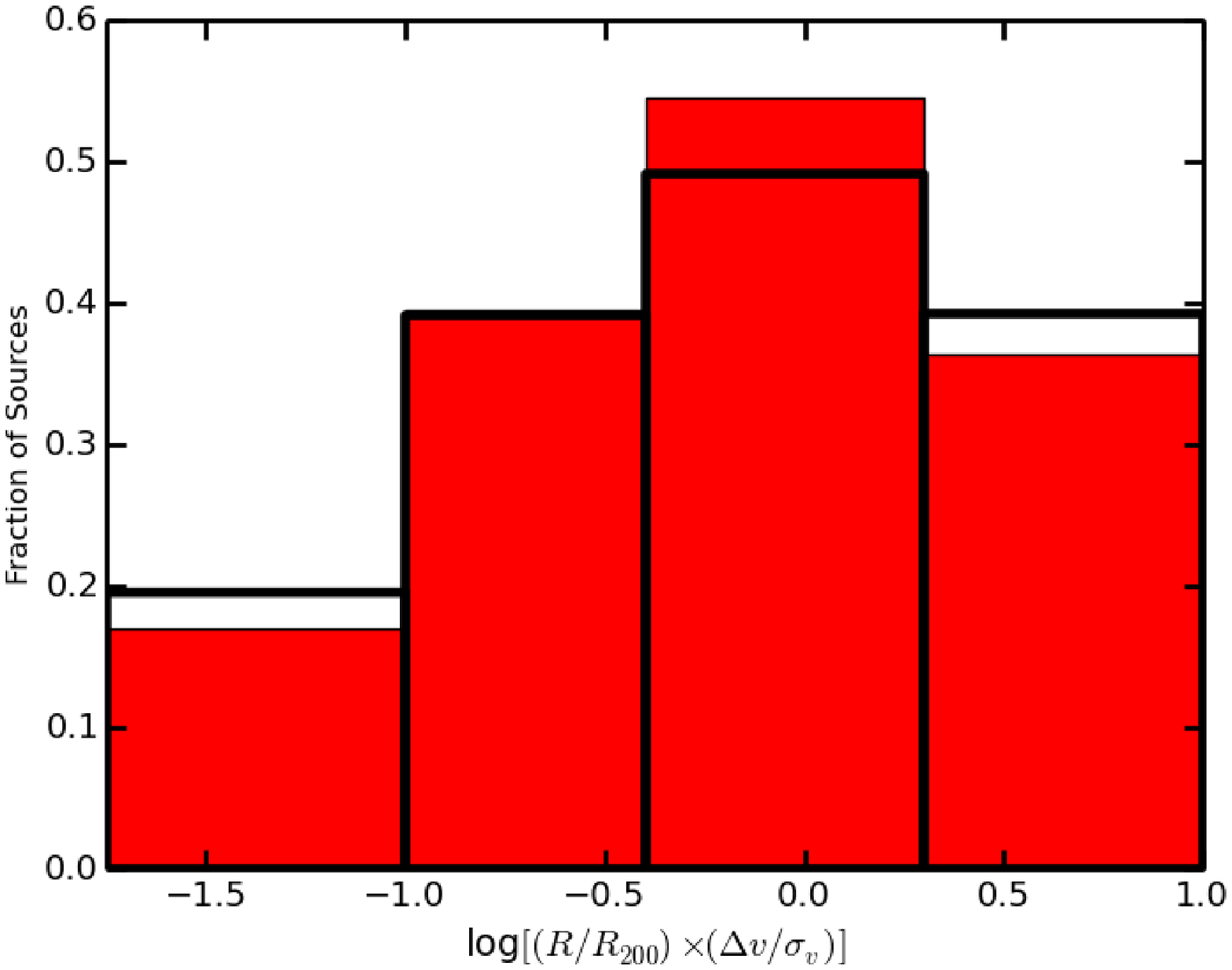}
\caption{
Plotted are histograms of $p=r/r_{200}\times \left|\Delta v\right|/\sigma_v$ for AGNs (filled bars) and all spectroscopically confirmed members of LSSs in our sample (unfilled bars). For more details on how $p$ is measured, see Section \ref{sec:spatdist}. The four bins are separated by the same lines of constant $p$ as plotted in Figure \ref{PSDfig}, at $p=0.1$, 0.4, and 2. 
}
\label{PSDhist}
\end{figure}

In Figure \ref{PSDfig}, lines corresponding to $p=0.1$, 0.4, and 2 are plotted. As in \citet{noble13}, galaxies with $p<0.1$ are likely in a dense cluster or group core and have had time to viriliaze, while those with $p>0.4$ (and, in our case, $<2$) are likely part of recently-accreted infalling populations. The region $0.1<p<0.4$ is an intermediate region. Because of our different scope (we include a substantial number of galaxies in our sample at larger spatial and/or velocity distances from the nearest group or cluster), we also include the caustic line at $p=2$. Galaxies with $p>2$ are likely not associated with any group or cluster. In Figure \ref{PSDhist}, we plot the fraction of galaxies in each of the four regions created with our three caustic lines. The distributions are shown for all spectroscopically confirmed galaxies, and for just the X-ray sources. We can see that the distributions look similar, which is confirmed by a KS test. This suggests that AGNs follow the overall distribution of galaxies with respect to groups and clusters. This implies that triggering mechanisms are at most weakly dependent on global environment, at least in relation to the four regimes probed here: dense cluster cores,  intermediate regions, infalling populations, and our `superfield' populations ($p>2$) not associated with any particular group or cluster. 

\subsection{AGN Host Galaxy Colors}
\label{sec:hgalcolan}

Figure \ref{CMDS} presents color-magnitude diagrams, which are described in Section \ref{TRSATBP}, for all LSSs. Confirmed AGN members of the LSSs are shown with red diamonds. While many AGN hosts lie on the red sequence, there are also a number just below it, in the green valley. Generally, the green valley is sparsely populated compared to the red sequence and blue cloud, and may be a transitional region where blue galaxies are evolving onto the red sequence \citep{faber07}. Previous work, including \citet{koc09b}, \citet{rum12} and a number of wide-field surveys \citep{sanchez04,nandra07,georg08,silver08}, has found an association between active galaxies and the green valley. While an association with this region could have a number of different implications, we examined color offsets of the AGN hosts from the red sequence to quantify this association.

\begin{figure}
\includegraphics[width=0.5\textwidth]{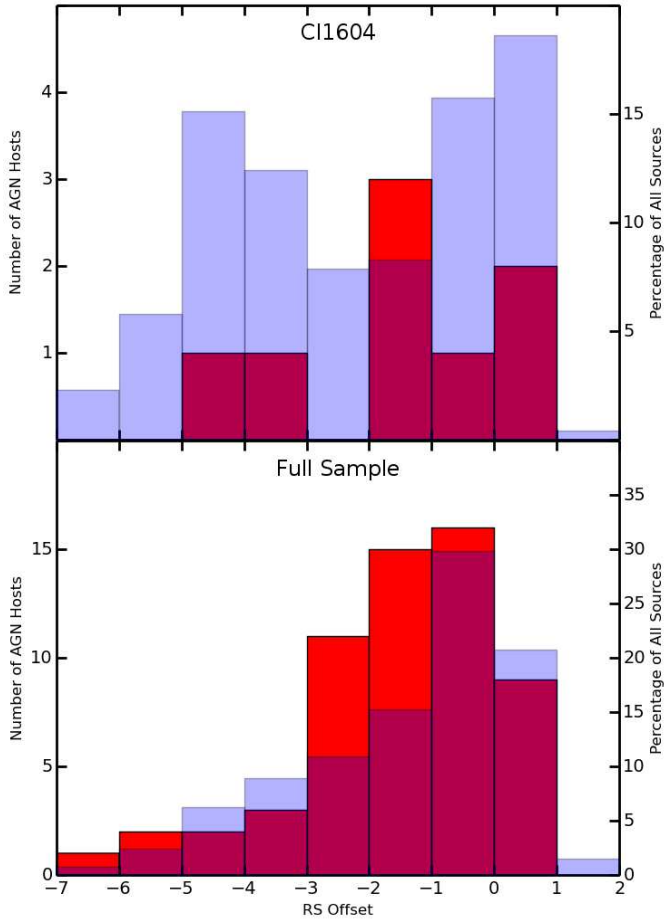}
\caption{
Histograms of the red sequence offsets are shown. The red sequence offset is defined as the distance below the center of the red sequence on a CMD, divided by half the width of the red sequence. The red histogram show the number of AGN hosts, while the lighter, blue, transparent histograms show the distributions for all spectroscopically confirmed member galaxies in our sample. The top plot shows the distributions for just the SC1604 LSS, while the bottom plot shows the distributions for the full sample (including SC1604). The red sequence, green valley, and blue cloud approximately correspond to red sequence offsets of $-1<\Delta$RS$\ <1$, $-3<\Delta$RS$\ <-1$, $\Delta$RS$\ <-3$ on these plots.
}
\label{RSOfig}
\end{figure}

To calculate red sequence offsets (RS offsets), we used the fits to the CMDs described in Section \ref{TRSATBP}. For a given galaxy, the RS offset is defined as the vertical distance above the center of the red sequence, divided by the half-width of the red sequence. The half-width is defined here as the vertical distance from the center of the red sequence to its boundary, so the red sequence ranges from an RS offset of -1.0 to 1.0. Histograms of RS offsets are shown in Figure \ref{RSOfig}. The top panel shows only SC1604, while the bottom panel shows all LSSs (including SC1604). For SC1604, the CMDs were constructed using ACS colors\footnote{Note that the ACS spatial coverage is smaller than the LFC coverage, meaning two of the SC1604 AGNs are not included in our color analyses. We do not expect the omission of these two AGNs to significantly affect our results, as these two excised AGNs did not have exceptional X-ray luminosities or LFC colors compared to the other SC1604 AGNs.}. Since ACS data were only available for SC1604, the other diagrams were constructed using our so-called supercolor parameterization of the $R$, $I$, and $Z$ colors, described in Section \ref{TRSATBP}. The RS offsets therefore use the ACS colors for SC1604 and the supercolors for all other LSSs. Because SC1604 used a different color scheme, and because it has a very well-defined blue cloud unlike many other members of our sample, we plot it by itself in the top panel of \ref{RSOfig}. The nature of the red sequence offsets allows us to easily compare the different color schemes, and the different LSSs with different red sequence fits. In the bottom panel of \ref{RSOfig}, we plot the RS offsets for all of our sample, including SC1604. For both plots, the distribution of AGN hosts only is shown with the darker red histograms. All other spectroscopically confirmed members of the LSSs are shown with the blue, lighter, transparent histograms. For the AGN hosts, the number in each bin are plotted, as shown on the left axes. For all LSS members, the percentage in each bin is shown on the right axes. The AGN histograms are scaled so that both the left and right y-axis labels apply. 

In the top plot, we can see that the overall galaxy population in SC1604 is bimodal: it has a well-formed blue cloud and red sequence, with associated peaks in the red sequence offset, with a trough in between, which represents the green valley. From this plot, the approximate area of the green valley is a red sequence offset between -3 and -1. In SC1604, 38\% of AGN hosts lie in this region, while only 19\% of the rest of the galaxies do. This is a marked difference, although the small number of AGNs mean it is not significant. 

In contrast, looking at our entire sample in the bottom plot, there is no bimodality in the overall galaxy distribution. This is likely because many of the LSSs in our sample do not have well-formed blue clouds, meaning we see no peak at lower red sequence offsets values as in the SC1604 distribution. However, AGNs still appear to be over-represented just below the red sequence. For SC1604, the green valley appeared to be between a red sequence offset of -3 and -1. If we consider this same region, we have 44\% of the AGN population and only 27\% of the overall population. While a KS test finds that the two distributions in the lower plot differ at less than a 90\% confidence level, the results are still suggestive of a link between the green valley region and AGN activity.

\subsection{Spectral Analysis of AGN Hosts}
\label{sec:AGNspec}

\begin{figure}
\includegraphics[width=0.5\textwidth]{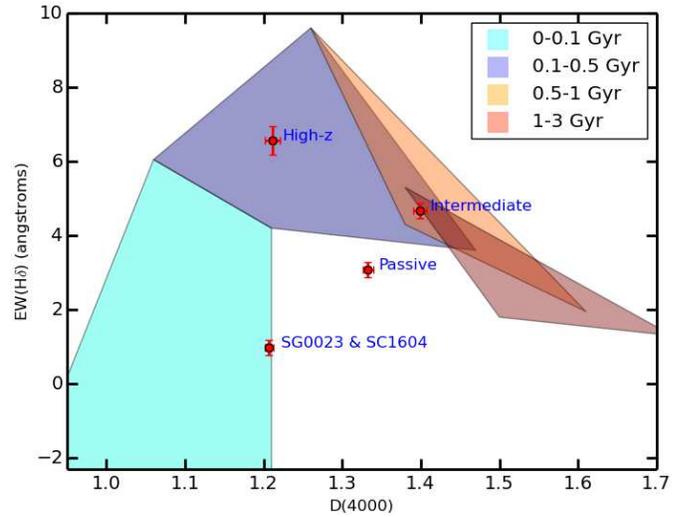}
\caption{
EW(\Hd) and \DFK\ measurements are plotted against each other for the composite spectra composed of the AGNs in each of the Passive, Intermediate, and High-$z$ sub-samples and SG0023 \& SC1604. Additionally, the evolutionary tracks for post-starburst galaxies based on a range of models and initial parameters are displayed. The legend notes the time since last starburst event for galaxies in each region. See Section \ref{sec:globspec} for more details. 
}
\label{agnspecfig}
\end{figure}

To investigate the connection between AGN activity and star formation, we created composite spectra of the AGN hosts, as in Section \ref{sec:globspec}. The spectra for all AGNs (with $L_x > 10^{42.5}$\ergs, see Section \ref{sec:xlum} and Section \ref{sec:xraycomp}) in each of the four sub-samples (Passive, Intermediate, SG0023+SC1604, and High-$z$) were coadded into a composite spectrum. These coadditions are necessary because of the small number of AGNs in most LSSs. Measurements of the spectral features of each of the four composite spectra are shown in Table \ref{AGNspectab}. In addition, the EW(H$\delta$) and \DFK\ values are plotted in Figure \ref{agnspecfig}, along with evolutionary tracks for post-starburst galaxies described in Section \ref{sec:globspec}. 

All four sub-samples have significant \OII\ emission. While \OII\ emission can be used as a proxy for ongoing star formation, in all cases, it is likely that most or a significant portion of this emission is from the AGNs themselves. In SC1604, six AGNs were analyzed by \citet{koc09a} using the Keck II Near-infrared Echelle Spectrograph \citep[NIRSPEC;][]{mclean98}. Five of these AGN hosts had $\left[N_2\right]$/H$\alpha$ ratios too high for normal star-forming galaxies, leaving AGNs as the most likely explanation for the source of the \OII\ emission. The average galaxies in the overall populations of the Passive and Intermediate sub-samples had low to middling star formation rates, which are highly unlikely to create the observed EW(\OII) measurements. The galaxy population in High-$z$ sub-sample was the bluest and had the most star formation, and we now find that their AGN hosts have the strongest \OII\ emission. This could be because their AGNs themselves have the strongest emission. In that case, we might expect to find more X-ray luminous AGNs in the High-$z$ sub-sample compared to SG0023 \& SC1604 which have much weaker \OII\ emission. However, as we discuss later in Section \ref{sec:xlum}, we find the opposite. Along with the blue colors and strong star formation in the overall galaxy population, it is likely that star formation is contributing to the high EW(\OII) measurements for the High-$z$ sub-sample. Even so, because of the likely strong contribution to the EW(\OII) measurements from AGNs in all cases, we cannot deduce much about the star formation in the AGN hosts based on this spectral feature. 

As described in Section \ref{sec:globspec}, the H$\delta$ line can be used as a proxy for recent star formation, as it tracks the A and B stellar population. While strong H$\delta$ absorption indicates recent star formation, weak absorption can indicate either an older stellar population where A and B stars have died out, or a very young population dominated by O stars. Analysis in conjunction with the 4000\AA\ break can break this degeneracy, as a larger \DFK\ measurement correlates with an older stellar population. If we examine Figure \ref{agnspecfig}, we see the average AGN host in SG0023 \& SC1604 has little H$\delta$ absorption and a low \DFK\ measurement. This places the average host in these LSSs near the beginning of the post-starburst evolutionary tracks, meaning a starburst event is either ongoing or has just ended. This is consistent with the relatively blue colors and star formation activity in the average galaxy in the overall populations, although the AGN hosts are much more active.

The AGN hosts in the High-$z$ sub-sample have a high average EW(\Hd) and a larger value of \DFK\ than for SG0023 \& SC1604. The average High-$z$ host has had a starburst within the last 100-500 Myr, according to these measurements. This is less time than the average galaxy in the overall population of these LSSs, although it is only about equal to Cl0849, which appears to be the most actively star-forming LSS in our sample. We see a similar trend for the Intermediate and Passive AGN hosts. In both cases the average AGN host in these sub-samples has a larger EW(\Hd) and a smaller \DFK. At this point on the evolutionary tracks, this indicates a smaller time since last starburst. 

We can conclude that, in all cases, the average AGN host has had less time since its last starburst event than the overlying galaxy population. It is also notable that, while the average High-$z$ galaxy has had less time since last starburst than the average galaxy in SG0023 or SC1604, the opposite is true of AGN hosts, with those in SG0023 \& SC1604 having the least average time since last starburst. This seems to be because of the exceptional activity level in SG0023 \& SC1604 AGN hosts. In the other three sub-samples, the average AGN host still had less time since last starburst than the average galaxy in the overall population, but the differences were moderate in comparison. This suggests that there is something exceptional about the AGN hosts in SG0023 \& SC1604.

\begin{table}
\caption{AGN Hosts' Average Spectral Properties}
\label{AGNspectab}
\begin{tabular}{lccc}
\footnotesize{Sub-Sample}
 & \footnotesize{EW(\OII)}
 & \footnotesize{EW(\Hd)}
 & \footnotesize{\DFK}\\
\midrule
Passive & \phantom{1}-7.1 $\pm$ 0.3 & 2.8 $\pm$ 0.2 & 1.30 $\pm$ 0.01\\
Intermediate & \phantom{1}-8.4 $\pm$ 0.4 & 5.3 $\pm$ 0.2 & 1.5 $\pm$ 0.01\\
SG0023+SC1604 & \phantom{1}-8.0 $\pm$ 0.2 & 1.0 $\pm$ 0.2 & 1.21 $\pm$ 0.01\\
High-$z$ & -12.2 $\pm$ 0.5 & 6.6 $\pm$ 0.4 & 1.21 $\pm$ 0.01\\
\bottomrule
\end{tabular}
\end{table}

\subsection{Close Kinematic Pairs}
\label{sec:CKP}

To investigate a connection between AGN triggering and galaxy interactions, we can locate close kinematic pairs (CKPs), which we define as two galaxies within 70 $h^{-1}$ kpc on the sky and within 350 km s$^{-1}$, typical values for identifying close pairs \citep[See, e.g.,][]{naz14,robo14,BB15,wis15,con16}, using our spectroscopic sample. While galaxies that have recently undergone a merger may not have a close kinematic pair, as the two galaxies may have coalesced, or they may still be too close together to resolve, galaxies that have recently had a more minor interaction should be more likely to lie in close proximity, as should galaxies that will soon undergo interactions. While this analysis cannot tell us which specific galaxies have recently merged or interacted, it can tell us which LSSs have galaxy interactions happening more frequently, since their galaxy populations should have more pairs of galaxies either about to undergo interactions or having recently undergone interactions that did not result in coalescence, and thus appearing as CKPs. In short, CKP fraction is a proxy here for how likely the average galaxy in a LSS is to have recently undergone a merger, even though we cannot tell if an individual AGN host has recently interacted.  

The percentages of AGNs and all spectroscopically confirmed galaxies in each sub-sample with CKPs are shown in Table \ref{CKPtab}. Although only a handful of AGN hosts in our sample have a CKP, SG0023 \& SC1604 have large fractions of AGN hosts with CKPs. At 35\% of the AGN population in these LSSs, this is over twice the percentage of AGNs with CKPs in any of the other three sub-samples. In fact, at 21\%, the fraction of CKPs in the overall galaxy population in SG0023 \& SC1604 is the highest out of our sample\footnote{Our photometric redshift catalogs predict that relatively few CKPs have been missed due to incompleteness, so this is unlikely to be the cause of the differences observed in Table \ref{CKPtab}.}, as well. Also, while the low number of AGNs in our LSSs means the fraction with CKPs has a high uncertainty, the combined fraction of AGNs in the Passive, Intermediate, and High-$z$ sub-samples is more precise, and is still only 11\%, considerably lower than for SG0023 \& SC1604.  

In addition to comparing between our sub-samples, we can also look at the field galaxies in our sample. For this analysis, we considered a field galaxy to be any outside the redshift bounds of our LSSs, but still within the overall redshift bounds of our sample ($0.65<z<1.28$). The fraction of CKPs in this sample is relatively low, as shown in Table \ref{CKPtab}, with only 9.5\% of the AGN population and 6.7\% of the overall population having CKPs. This is unsurprising, since the less dense field environment should be less conducive towards galaxy interactions. And, again, while our sample is small, the CKP fraction is not significantly higher for the combined AGN population of the Passive, Intermediate, and High-$z$ sub-samples than for the field galaxies. For the overall populations, the Passive and High-$z$ subsamples also do not have CKP fractions higher than that for the overall field population at the 3$\sigma$ level. 

On the other hand, the relatively large number of AGN hosts with CKPs in SG0023 \& SC1604 are likely related to the large fraction of CKPs in their overall populations. Notably, the AGN hosts in these LSSs are alone in having a CKP fraction significantly above zero. While small number statistics make the comparison to the other LSS AGN populations somewhat uncertain, these results suggest that the AGN population in SG0023 \& SC1604 is influenced by a higher rate of galaxy interactions among their hosts.

\begin{table}
\caption{Close Kinematic Pairs}
\label{CKPtab}
\begin{tabular}{llll}
\footnotesize{Sub-Sample}
 & \footnotesize{Num. of}
 & \footnotesize{Perc. of}
 & \footnotesize{Perc. of}\\
 \footnotesize{ }
& \footnotesize{AGNs}
& \footnotesize{AGNs}
& \footnotesize{all Gal.}\\
 \footnotesize{ }
& \footnotesize{with CKP}
& \footnotesize{with CKP}
& \footnotesize{with CKP}\\
\midrule
Passive & 2 & 11 $\pm$ 7 & 11 $\pm$ 2\\
Intermediate & 1 & 8 $\pm$ 11 & 15 $\pm$ 2\\
SG0023+SC1604 & 6 & 35 $\pm$ 10 & 21 $\pm$ 2\\
High-$z$ & 2 & 14 $\pm$ 7 & 6 $\pm$ 2\\
Field & 8 & 9.5 $\pm$ 3 & 6.7 $\pm$ 0.6\\
\bottomrule
\end{tabular}
\end{table}

\subsection{X-ray Luminosity}
\label{sec:xlum}

In this section, we examine the differences in the X-ray luminosities of the AGN populations in our sample. We calculate rest-frame luminosities for X-ray point sources with
known redshifts, using a power law
spectral model with a photon index of $\gamma=1.4$,
described in Section \ref{sec:red}. 

\begin{figure*}
\includegraphics[width=\textwidth]{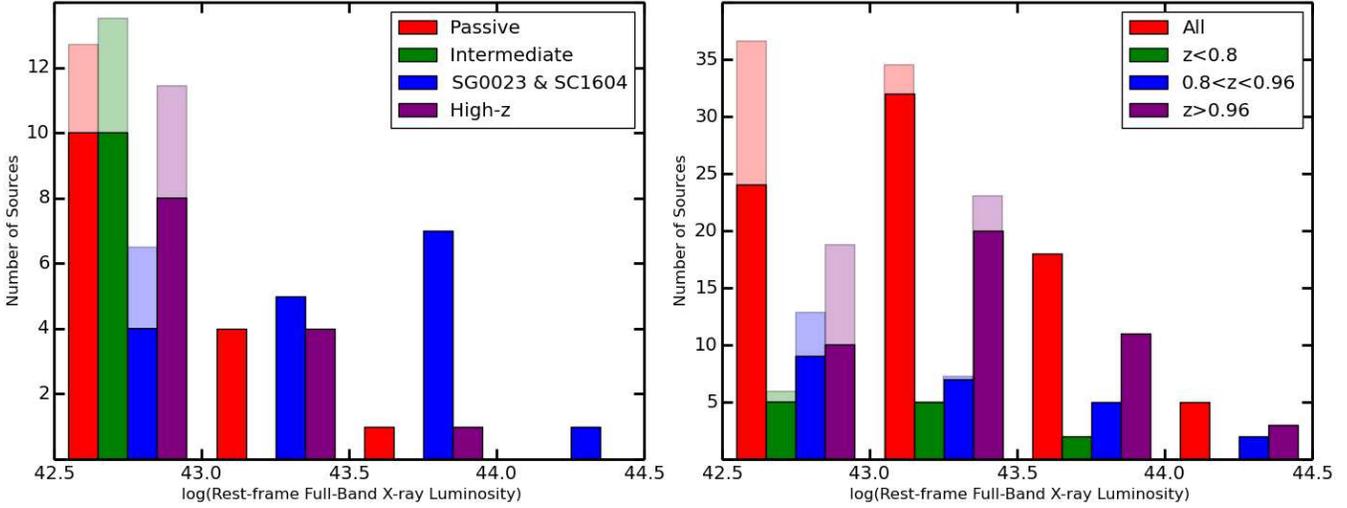}
\caption{
Histograms of the rest-frame, full-band X-ray luminosities of AGNs in our sample are shown. The x-axis is in log-space, representing log$\left(L_x\right)$. In the left panel, luminosities are shown for all AGNs within the LSSs in our sample, sub-sampled as described in Section \ref{sec:SGC}. In the right panel, the luminosities for field galaxies in our sample are shown. See Section \ref{sec:CKP} for a description of their selection. In both plots, corrections for the X-ray luminosity limits are shown with lighter bars. See Section \ref{sec:xraycomp} for details about these completeness calculations.
}
\label{XrayLumfig}
\end{figure*}

X-ray luminosities for all member AGNs are listed in Table \ref{AGNtab}, and the full-band luminosities are displayed in the left panel of Figure \ref{XrayLumfig}. Objects with luminosities below 10$^{42.5}$ ergs s$^{-1}$ were cut from the sample because of low completeness in this regime, and the lighter bars in the histogram represent corrections for completeness. See the Section \ref{sec:xraycomp} for more details. The luminosity distributions for the Passive, Intermediate, and High-$z$ sub-samples all peak in the lowest luminosity bin, falling off sharply towards the highest bin, especially the Intermediate sub-sample, which only has AGNs in the lowest luminosity bin. The luminosity distribution for SG0023 \& SC1604 is noticeably different from that of the rest of the AGNs, peaking at high luminosity in contrast to the other sub-samples. These LSSs have a disproportionate number of AGNs with high luminosities, containing eight out of the ten AGNs with luminosities above 10$^{43.5}$ ergs s$^{-1}$. The difference between the distributions is confirmed by a KS-test at the 3$\sigma$ level, when comparing the SG0023 \& SC1604 to the distribution of all other AGNs. Although correcting for our X-ray luminosity limits, as discussed in Section \ref{sec:xraycomp}, does slightly alleviate the discrepancy, it does not explain the discrepancy. With these corrections, a KS-test finds the SG0023 \& SC1604 luminosity to be different from the other sub-samples at the $\ga$97\% level. 

These discrepancies could also be caused by spectroscopic incompleteness. It is possible that bluer, more luminous AGN hosts were missed if redder galaxies evaluated to be more likely to reside in the LSSs were preferentially targeted in our slit masks instead. To estimate the number of galaxies complete spectroscopy would add, we use our photometric redshift catalogs, although we can only perform the analysis for a subset of our sample as photometric redshifts have only been calculated for SG0023, RXJ0910, SC1324, SC1604, RXJ1716, RXJ1757, and RXJ1821 so far. For each X-ray point source matched to a galaxy in our photometric catalog without a high quality redshift in the fields of view of these {\it Chandra} observations, we estimated the probability that source was within the redshift range of the respective LSS (See Section \ref{sec:photzcomp} for more details). Ignoring objects with less than a 5\% chance of membership, we estimate that 0.2, 1.7, 0, 0, 0.8, 0, and 0.2 sources would be added in SG0023, RXJ0910, SC1324, SC1604, RXJ1716, RXJ1757, and RXJ1821, respectively, with complete spectroscopy. Because these estimates are small (increasing our sample size, on average, by $\sim$10\%), it is unlikely that spectroscopic completeness is causing any discrepancies between the luminosity distributions of our sub-samples. 


In the right panel of Figure \ref{XrayLumfig}, we plot the X-ray luminosities for the field galaxies in our sample, as defined in Section \ref{sec:CKP}. Again, the lighter bars in the histogram represent corrections for completeness. These AGNs are binned by redshift to roughly correspond to the redshifts of the sub-samples we made for our LSS AGNs. The redshift range $0.8<z<0.96$ roughly corresponds to the combined redshift range for SG0023 \& SC1604. Most of the galaxies in the Passive sub-sample are at $z<0.8$, while only the galaxies in the High-$z$ sub-sample have redshifts greater than 0.96. Both similarities and differences can be seen between the field and LSS AGN luminosity distributions. The $z<0.96$ field AGN luminosity distributions, like all of those for the LSS AGNs except SG0023 \& SC1604, peak in the lowest luminosity bin and fall off those higher luminosities, although they do not appear to fall off as quickly. The $z>0.96$ field AGN luminosity distribution peaks in the second lowest luminosity bin, unique among populations studied here, and this does not appear to be a product of lower luminosity limits at higher redshifts. A KS test finds that the $z>0.96$ field distribution differs from the $z<0.8$ and $0.8<z<0.96$ field AGN distributions at the 94\% and 89\% levels, approximately. The $z>0.96$ field AGN luminosity distribution is also found to differ from the Passive, Intermediate, and High-$z$ sub-samples at the 3$\sigma$ level by a KS test. KS tests find that the Intermediate distribution is likely to be different from all field distributions at the 99\% level, in fact, although this is probably related to the peculiar absence of $L_x > 10^{43}$ \ergs AGNs in these LSSs. None of the other sub-sample luminosity distributions are found to vary from those of the $z<0.96$ field at a significant level. Additionally, since the SG0023 \& SC1604 distribution was found to be different from the other LSS luminosity distributions, but was not found to be significantly different from any of the field distributions, this suggests that the field AGN luminosity distributions are partway between those of SG0023 \& SC1604 and the other sub-samples. 

These differences between the field AGNs and those in the LSSs could be explained by higher reservoirs of gas in the field AGNs, which tend to be bluer. This would tend to create more luminous AGNs, which would explain the milder falloff toward higher luminosities when compared to the Passive, Intermediate, and High-$z$ sub-samples. This could also explain the higher luminosity of the $z>0.96$ field AGNs, as the blue fraction for galaxies tends to increase with redshift. Because of the low CKP fractions in the field populations, even at high redshift, it is unlikely that the mechanisms that create high luminosity AGNs in SG0023 \& SC1604 are doing the same in the field. The similar shape of the LSS AGNs and $z<0.96$ field luminosity distributions (excluding SG0023 \& SC1604) could also indicate that similar process are triggering AGNs in both samples. The marked difference between the SG0023 \& SC1604 luminosity distributions and those of all other AGNs suggests different triggering mechanisms may be taking precedence, which was also indicated by the CKP fractions.

\subsubsection{Relation to Host Colors and Spectroscopy}
\label{sec:RSOvsLum}

\begin{figure*}
\includegraphics[width=\textwidth]{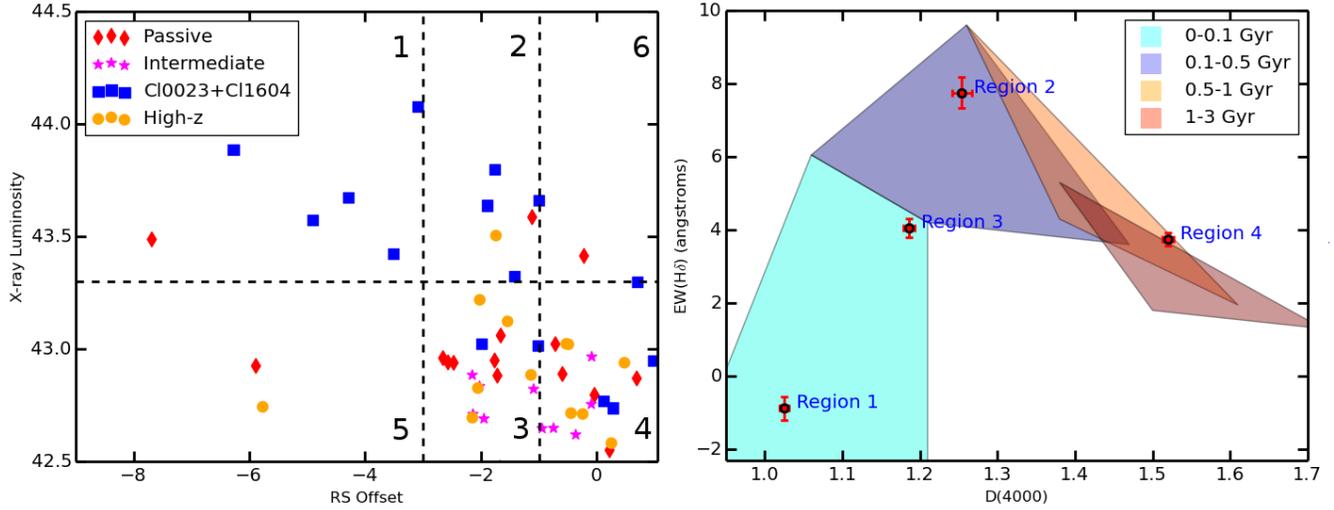}
\caption{
{\it Left}: Offsets from the red sequence for each AGN host are plotted against the AGN's full-band, rest-frame X-ray luminosity. See Section \ref{sec:hgalcolan} for an explanation of the calculation of the red sequence offsets and Section \ref{sec:xlum} for discussion of X-ray luminosities. The dashed lines separate the galaxies and their AGNs into six regions. Regions 4 and 6 make up the red sequence, Regions 2 and 3 represent the green valley, and Regions 1 and 5 are the blue cloud. {\it Right}: EW(\Hd) and \DFK measurements are plotted for the composite spectra for the AGNs in Regions 1, 2, 3, and 4 of the left panel. Additionally, evolutionary tracks for post-starburst galaxies based on a range of models are displayed. The legend notes the time since last starburst event for galaxies in each region. See Section \ref{sec:globspec} for more details. 
}
\label{RSOvsLumfig}
\end{figure*}

In the left panel of Figure \ref{RSOvsLumfig}, we plot the full-band, rest-frame X-ray luminosities versus the red sequence offsets, as described in Section \ref{sec:hgalcolan}. We have divided the plot into six regions based on X-ray luminosity and red sequence offsets. The x-axis is broken up into three regions, delineating, from left to right, the blue cloud, green valley, and red sequence. On the y-axis, we break up the plot into two sections based on luminosity, separating AGNs at $L_x=10^{43.3}$\ erg s$^{-1}$. This boundary was chosen in \citet{rum12} and kept for ease of comparison to that work. Almost all of the red sequence AGNs lie below this line and most of the blue cloud AGNs lie above it. This result is unsurprising, since red sequence AGN hosts should have smaller reservoirs of cold gas for fuel, and should tend to house lower luminosity AGNs, while the opposite would be expected in the blue cloud. However, the large range of AGN luminosities in the green valley (red sequence offsets between -3 and -1) is notable. While we found previously in \citet{rum12} that no AGNs inhabited Regions 5 or 6, which correspond to low luminosity blue AGN hosts and high luminosity hosts on the red sequence, respectively, we do have a few such objects in our current sample.

We can examine these regions in further detail by looking at their hosts' spectra. As in Sections \ref{sec:globspec} and \ref{sec:AGNspec}, we coadded the spectra to make composite measurements of the hosts' properties. AGNs in each of Regions 1, 2, 3, and 4 were coadded. The small number of AGNs in Regions 5 and 6 precluded coaddition. To examine the evolutionary states of the AGNs in each region, we measured EW(H$\delta$) and \DFK\ for each composite spectrum. The results are plotted in the right panel of Figure \ref{RSOvsLumfig}. Also plotted are evolutionary tracks for galaxies for varying time since last starburst. 

The low values of EW(H$\delta$) and \DFK\ indicate that the Region 1 AGNs, which are in the blue cloud, have ongoing or very recent starburst events, or concurrent star formation. In contrast, the Region 4 AGNs, which are on the red sequence, have had the most time, on average, since the last starburst event. Intermediate to these are the Region 2 and 3 AGNs. Both have has less time than the Region 4 AGNs and more time than the Region 1 AGNs since last starburst, on average. However, the fainter Region 3 AGNs appear to have had less time since starbursting. This conflicts with our previous findings presented in \citet{rum12}, which had a smaller sample of 27 AGNs, and where we found that the average Region 2 AGN host had less time since last starburst event than for Region 3. The change in the temporal order of the AGNs on the green valley, after increasing sample size, may indicate that green valley AGNs across all luminosities have similar properties, or could be a sign of an underlying process that is correlated with X-ray luminosity. 

We can conclude, though, that AGNs across all parts of the CMD have had a starburst, on average, within the last Gyr. There appears to be a close connection between starbursts and AGN activity, particularly for blue cloud and green valley galaxies, especially since the EW(H$\delta$) values are too large to be attributed to normal star formation. In fact, there seems to be an evolutionary correlation to the host galaxy colors, as blue cloud, green valley, and red sequence AGN hosts have had successively more time since the last starburst event, on average.

\section{Discussion}
\label{sec:dis}

We studied AGNs and their host galaxies in 12 large-scale structures in the ORELSE survey. Using {\it Chandra} ACIS-I and ACIS-S observations, we identified X-ray point sources in our sample, then matched them to our extensive spectroscopic catalogs using a maximum likelihood technique described in Section \ref{sec:OM}. Our DEIMOS spectroscopic observations have targeted $\ga 1000$ objects per field, on average. From this data set, we have obtained redshifts for 288 X-ray point sources, locating 61 AGNs within the LSSs in our sample. 

Our sample spans a wide-range of environments, from isolated clusters to superclusters, and even includes a system of five merging groups. In addition, our redshift range is from 0.65 to 1.28. These qualities provide an excellent opportunity to study AGN activity and triggering mechanisms. Many of the LSSs in our sample had only a few AGNs, however. Without sufficient numbers to analyze the LSSs alone, we separated our sample into four sub-samples, based on the properties of the LSSs. 

To determine how to subdivide our sample, we measured the average spectral properties of each LSS. We accomplished this through composite spectra created by coadding all spectroscopically confirmed members of each LSS. As described in Section \ref{sec:globspec}, we used the \OII\ and \Hd\ lines and the 4000\AA\ break to probe star formation histories. The \OII\ line is a proxy for very recent or ongoing star formation, caused by O stars dominating the spectrum. The \Hd\ line probes recent star formation, characterized by an older populations of A and B stars, which disappear within a Gyr. The 4000 \AA\ break is a proxy for the average age of the stellar population. With these metrics, we can identify five LSSs that make up what we refer to as the Passive sub-sample. These are mostly isolated clusters at lower redshifts. The average member galaxies have the smallest values of EW(\OII) and EW(\Hd), and the largest average \DFK, indicating an older, quiescent population of galaxies. Additionally, they have well-defined red sequences and low blue fractions, which are characteristic of quiescent clusters. This sub-sample stands in stark contrast to Cl0849, RXJ0910, and RXJ1053, the highest redshift LSSs in our sample, and which we refer to as the High-$z$ sub-sample. The average member galaxies of these LSSs have the largest values of EW(\OII) and EW(\Hd), and the lowest average \DFK, indicating they are the most active LSS galaxy populations in our sample, substantiated by the large blue fractions. In \citet{rum12}, we found SG0023 and SC1604 to be the most active LSSs. While their spectral properties put them at a lower level of average star formation than the High-$z$ sub-sample, and they are not quite as blue, they are still more active than the rest of our sample, so we placed them into their own sub-sample. We are left with SC1324 and RXJ1716. These two LSSs straddle the line between a quiescent and star-forming average galactic population, in between the Passive sub-sample and SG0023 \& SC1604. As such, we christened them the Intermediate Sub-Sample. 

As we explore triggering mechanisms, we first discuss the colors of the AGN host galaxies, which provide valuable information on the properties of the host. On a CMD, galaxies tend to form bimodal distributions, where redder, more quiescent galaxies congregate on the red sequence, and bluer, more active galaxies form the blue cloud. In between is the sparsely populated green valley, which may be a transitional region where blue galaxies are rapidly evolving onto the red sequence after quenching star formation \citep{faber07}. Some studies have found a correlation between AGN activity and location in the green valley \citep{sanchez04,nandra07,georg08,silver08, koc09b, rum12}, although some studies using stellar mass-matched galaxy samples do not see the correlation \citep[See, e.g.,][]{silver09b,xue10}. To examine if our AGN hosts cluster in certain regions on a CMD, we analyzed their offsets from the red sequence. After performing fits to the red sequence for each LSS in our sample, we calculated red sequence offsets for each galaxy as its distance below the center of the red sequence, divided by the half-width of the red sequence. This allowed us to compare colors between our different LSSs. We found that AGN hosts were indeed over-represented in the green valley: 44\% of the AGN population was in this region compared to only 27\% of the overall population. However, this result is not particularly strong, as a KS test estimated the significance of the difference at less than 90\%. When accurate stellar masses are available for our entire sample, we can determine if this association is primarily related to higher mass hosts by comparing to a stellar mass-matched sample. 

An association with the green valley on a CMD could provide evidence for a number of different triggering mechanisms. For major mergers an association would be expected as AGN activity would be coeval with the aftereffects of quenching star formation in blue cloud galaxies. For scenarios involving rejuvenation of red sequence galaxies, AGNs would also tend to lie in the green valley as the host evolves blueward temporarily. Since different mechanisms could be more prominent in different global environmental regimes (e.g., galaxy mergers may be more likely to take place in the outskirts of clusters, where infalling populations lie), we also examined the spatial distributions of our AGN hosts. To compare AGN hosts near clusters and sub-samples of varying sizes, we employed the phase space metric $p=r/r_{200}\times \left|\Delta v\right|/\sigma_v$. See Section \ref{sec:spatdist} for more details. We separated galaxies into four environmental bins, using $p=0.1$, 0.4, and 2 as the boundaries. As shown in Figure \ref{PSDhist}, the phase space distribution of the AGN hosts is very similar to that of the overall galaxy population, and a KS test did not provide any significant evidence of difference between them. This result implies that the environment of AGN hosts, whether they are in dense cluster cores, on their outskirts, or far from any sub-samples or clusters, has no meaningful effect on their nuclear activity, at least when considering X-ray bright objects. This could mean that major mergers are not the dominant triggering mechanism in our sample, as we would have expected an association with intermediate phase space regimes, although our results in this regard are only suggestive.

Similar to how we examined the average spectral properties of the overall galaxy populations, we created composite spectra of the AGN hosts in each sub-sample. Using the \Hd\ absorption and 4000\AA\ break we found that, for all four sub-samples, the average AGN host has had less time since the last starburst event than the average galaxy in its overall population. This shows a strong correlation between AGN activity and starburst events, across our entire redshift range and LSSs at all levels of star formation activity. This correlation could also be related to the association we found between the green valley and AGN activity: if AGN hosts are more likely to be in the green valley than to be quiescent red sequence galaxies, we may expect them to also have had less time on average since the last starburst. However, the difference in time since last starburst for the SG0023 \& SC1604 populations exceeded the other sub-samples by a substantial margin. While the overall galaxy populations of these LSSs were not the most actively star-forming, a position held by the High-$z$ sub-sample, the average AGN host in these LSSs was the most active by far, with the average age since last starburst $\la 100$\ Myr. The starbursting, or nearly starbursting, AGN hosts in SG0023 \& SC1604 appear exceptional, and a similar picture is painted by their X-ray luminosity distribution. 

Examining the rest-frame X-ray luminosities of the AGNs in our sample, we found some stark contrasts between the different sub-samples. The Passive, Intermediate, and High-$z$ sub-samples all had most of their AGNs at lower luminosities, with only two with luminosities above 10$^{43.5}$ \ergs. Looking at Figure \ref{XrayLumfig}, we can see that all three distributions peak in the lowest energy bin. In contrast, the AGNs SG0023 \& SC1604 peak at higher energies. With eight AGNs with luminosities above 10$^{43.5}$ \ergs, the population is disproportionately luminous. Studies examining cluster AGNs have found that both the AGN fraction and the high-luminosity AGN fraction increase with increasing redshift, up to at least $z\sim1.5$ \citep[See, e.g.,][]{mart09,mart13,alberts16}. The disproportionately luminous SG0023 \& SC1604 AGN population runs counter to this finding, as we do not find a similarly luminous AGN population in the High-$z$ sub-sample. This implies something different is occurring in SG0023 \& SC1604 than the rest of our sample, which cannot be explained by more star formation activity or a younger overall galaxy population, as the High-$z$ sub-sample has these properties, but does not have a similar luminosity distribution. 

Our analysis of the spectral properties and X-ray luminosity distributions of the AGN hosts suggests some different mechanism is at play in the SG0023 \& SC1604 LSSs. One potential explanation could be an increased role of galaxy mergers or tidal interactions in AGN triggering. We examined the possibility by estimating the incidence of galaxy interactions through the proxy of close kinematic pairs. While a close kinematic pair does not necessarily pick out galaxies undergoing mergers or tidal interactions, a population with higher rates of mergers and tidal interactions should have more galaxies located close to each other, so the CKP fraction should be applicable as a proxy of galaxy interaction rate. We define a CKP as two galaxies within 70 $h^{-1}$ kpc on the sky and 350 km s$^{-1}$ in velocity space. We found that the SG0023 \& SC1604 AGN hosts had the highest fraction with CKPs, although we had low precision because of the small number of AGN hosts with CKPs. When looking at the overall galaxy populations, that of SG0023 \& SC1604 also had the highest fraction of CKPs. The high CKP fractions are likely indicative of a higher degree of galaxy interactions in these LSSs, and could explain the exceptional spectral and X-ray properties of the AGN hosts. This is supported by morphological analysis of SC1604 carried out by \citet{koc09b}, who found that two-thirds of AGNs examined had signs of recent or pending mergers or tidal interactions. 

As a control test, we also examined the properties of field galaxies in our sample. We defined field galaxies as those that, in each observation, that were outside the redshift bounds of the LSS observed, but still within the overall redshift range of the sample ($0.65<z<1.28$). We found that both the AGN hosts and overall galaxy populations of the field galaxies had low CKP fractions, which is expected given that field galaxies are in less dense environments on average, thus having fewer galaxy interactions. The X-ray luminosity distribution of the field AGNs at $z<0.96$ peaked in the lowest energy bin, similarly to the Passive, Intermediate, and High-$z$ sub-samples. The field AGNs at $z>0.96$ peaked at a slightly higher energy, but still at lower energies than the for SG0023 \& SC1604. The higher luminosities of the higher redshift field AGNs could be explained by larger available gas reservoirs to fuel AGNs in the bluer galaxies at these redshifts. Note, though, that SG0023 \& SC1604 combined lie within the redshift range $0.8<z<0.96$, and the X-ray luminosity distributions of these lower redshift field AGNs were more similar to the other sub-samples than to the SG0023 \& SC1604 AGNs. These results imply that similar processes are triggering AGNs in the field galaxies and in the Passive, Intermediate, and High-$z$ sub-samples, while some different mechanism is substantially more prominent for those in SG0023 \& SC1604. The CKP fractions and the expected low merger rate in the field populations provide evidence that the galaxies in SG0023 \& SC1604 have more mergers or tidal interactions occurring, which are triggering AGN activity. 

The X-ray luminosities of the AGNs in our sample, in conjunction with the red sequence offsets, provide further information on triggering and evolutionary processes. When these properties are plotted against each other, as in Figure \ref{RSOvsLumfig}, a correlation can be seen. Those in the blue cloud tend to be more X-ray luminous, while those on the red sequence are overwhelmingly less luminous. There is a wide range of X-ray luminosities for AGNs hosted by green valley galaxies. The difference between the AGNs in blue and red hosts could be explained by larger gas reservoirs available in bluer galaxies. If the green valley is a transition region where blue cloud galaxies are rapidly evolving onto the red sequence, and this transition is related to AGN activity, we would expect these color-luminosity trends as well. AGNs triggered in blue cloud galaxies would start out at their most luminous. As star formation is quenched, the galaxy's color evolves towards red, at the same time as a drop in AGN luminosity as it loses its fuel source. 

In the left panel of Figure \ref{RSOvsLumfig}, we have demarcated Regions 1, 2, 3, and 4, which correspond to blue galaxies with bright AGNs, green galaxies with bright AGNs, green galaxies with faint AGNs, and red sequence galaxies with faint AGNs, respectively. If we constructed composite spectra of the AGNs in each region, we might expect the EW(\Hd) and \DFK\ measurements to paint a picture where AGN hosts have had more time since last starburst as you look sequentially at Regions 1, 2, 3, and then 4. This is what we found in \citet{rum12}, where our sample consisted of just 27 AGNs in SG0023, SC1324, SC1604, RXJ1757, and RXJ1821, compared to 61 now. When we perform this same exercise with our current sample, we find again that Region 1 AGNs have had the least time since last starburst and Region 4 AGNs the most time, but the situation has been reversed for those on the green valley. As shown in Figure \ref{RSOvsLumfig}, Region 3 AGNs, on average, appear to have had less time since last starburst since the more X-ray luminous Region 2 AGNs. This cannot be explained due to uncertainties on the measurements. This implies something more complicated is occurring than a simple scenario wherein all AGNs and their hosts begin in Region 1 with their triggering and evolve into Region 4 over time. 

Of note, SG0023 \& SC1604 are over-represented in Region 1. After our analysis in \citet{rum12}, we expected to add more objects to this region as our sample size increased\footnote{Note that all new observations in this study were from outside SG0023 \& SC1604.}. However, despite the number of AGNs in our sample rising from 27 to 61, we added only one X-ray luminous, blue cloud AGN. Similarly, we added only two X-ray luminous, green valley AGNs. We disproportionately added AGNs in Regions 3 and 4, leaving Regions 1 and 2 dominated by SG0023 \& SC1604. This trend seems to indicate that SG0023 \& SC1604 are exceptional when compared to the general sample. We have already established that these LSSs seem to have an elevated level of galaxy interactions, which likely affects the AGN activity within them. Our new results could be explained if two AGN evolutionary scenarios were occurring. First, merger-induced AGNs would tend to trigger in Region 1, and would evolve sequentially through Regions 2 and 3 and end up in Region 4. A second process would tend to create less luminous AGNs, which would be triggered in Region 3 and evolve into Region 4. While there is still a population of merger-induced AGNs that fit on the track moving through Regions 1, 2, 3, and 4, the second class of AGNs would lead to a decrease in the average time since last starburst in Region 3, explaining this new result. 

The exact nature of this second triggering mechanism is not something our data can fully illuminate. A number of possible mechanisms for AGN activation have been proposed as alternatives to major galaxy mergers, such as minor mergers or tidal interactions \citep{men01}, disk or bar instabilities, or recycling of stellar material \citep{CO07}, and some studies have proposed gradual ramp-downs of accretion and luminosity to explain low-luminosity AGNs \citep{croton06,mc07,shen07}. If minor mergers were the second prominent trigger, we would expect to see increased activity for this AGN mode in the LSSs with high CKP fractions, as increased galaxy interactions should create both more major and minor mergers. As such, our data do not prefer this possibility, since the second mechanism is not exclusive to SG0023 \& SC1604. Moving beyond minor mergers, a detailed morphological study needs to be performed, requiring high resolution, multi-band data across the entire sample, to search for galaxies with signs of recent mergers or interactions to definitively rule on the issue, and further substantiate our results here, as well as to search for blue galaxy cores that would be signatures of some of these mechanisms. Alternatively, precise measurements of the local density for every AGN host, and galaxy in the overall population, could shed more light on where galaxy mergers and interactions are most likely to occur, and we plan to use this method to study the ORELSE sample, as in Lemaux et al. (2016, in prep.). Using this method, we will look for specific differences between the AGN hosts and the underlying galaxy populations in SG0023 \& SC1604, as well as relative to the rest of the LSSs. 

Other AGN studies have found results both similar and differing from ours, although this may be unsurprising given the conflicting evidence on AGN triggering mechanisms. \cite{alberts16}, for example, studied AGNs in $\sim$250 galaxy clusters at $0.5<z<2$ from the IRAC Shallow Cluster Survey and the IRAC Distant Cluster Survey. They found an excess AGN fraction at $z>1$ compared to lower redshifts, with the fraction exceeding the field at $z>1.5$. They argue this is evidence for the increasing prevalence of merger-induced AGN activity at higher redshifts, but this is not reflected in our dataset. It is possible this is due to small number statistics, since we only studied three LSSs at higher redshifts than SG0023 \& SC1604. With a larger $z>1$ sample, we may also find more LSSs like SG0023 \& SC1604. \citet{eh15} also find evidence for merger-induced AGN activity, analyzing $>11000$ {\it Chandra}-selected AGNs in and behind 135 massive ($\ga 10^{14} M_{\odot}$) galaxy clusters at $0.2<z<0.9$. They find a spatial dependence on AGN fraction, with an excess in the central regions of clusters, a result not reflected in our data. Additionally, AGN abundance in their dataset scales with the parent cluster mass as $M_{500}^{-1.2}$, showing a suppression at higher masses. This is similar to the galaxy merger rate, suggesting a connection, compounded by a higher incidence of disturbed morphologies among AGN hosts. \citet{ros15} used Cosmic Assembly Near Infrared Extragalactic Legacy Survey (CANDELS) HST imaging and far-infrared imaging from the PEP+GOODS-{\it Herschel} survey to study $>$100 X-ray selected AGNs in two {\it Chandra} Deep Fields. Similarly to the previous studies, they find a higher incidence of clumpy morphologies suggestive of recent mergers among AGN hosts compared to inactive galaxies at $z\sim1$, although the connection weakens towards $z\sim2$. 

Providing evidence against mergers as AGN triggers, \citet{koul14} find no suppression of AGN activity at high densities relative to the field, examining X-ray-selected AGNs in 33 galaxy clusters in the XMM-{\it Newton} Large Scale Structure Sample at $0.14<z<1.05$. They do use a lower mass sample of galaxy clusters than \citet{eh15}, which includes only $\ga 10^{14} M_{\odot}$ clusters. The lack of correlation between our phase space metric, $p=r_{norm}\times v_{norm}$, and AGN fraction could similarly result from the lower masses of galaxy clusters and groups relative to \citet{eh15}. Additionally, \citet{HI16} analyzed SDSS-DR7 spectra of 385 isolated pairs of galaxies ($\Delta v\le1200$\kms, $r\le100$ kpc). When compared to the the 513 isolated galaxies studied by \citet{HI13}, they find similar AGN fractions, albeit at low luminosities ($L_{H\alpha}\sim2\times10^{40}$ \ergs). This result does leave open the possibility that major mergers trigger high-luminosity AGNs and is consistent with our findings that low-luminosity AGNs are not triggered by minor mergers or interactions. \citet{koc12} examine HST/WFC imaging from CANDELS of 72 moderate luminosity ($L_X\sim10^{42-44}$ \ergs) AGN hosts at $1.5<z<2$, selected using the 4 Ms {\it Chandra} Deep Field South. Similarly, they find both AGN hosts and non-active galaxies have similar fractions with disturbed morphologies, which is consistent with the $z\sim2$ sub-sample of \citet{ros15}. This may continue the trend observed in our sample, where our $z>1$ LSSs had CKP fractions and X-ray luminosity distributions similar to the field, but does not preclude major mergers playing an important role at lower redshift. 

One of the most unique aspects of our sample relative to other studies is the approximately binary nature of the possibly merger-induced, high-luminosity AGNs. Rather than merely correlating with an environmental or host property such as CKP fraction, these high-luminosity AGNs are concentrated almost entirely within two LSSs. This suggests the merger AGN mode is almost entirely suppressed elsewhere. While our results may be influenced by a small sample size, in that only two LSSs appeared to show a preference for merger-induced AGN activity, they may suggest that the properties of some LSSs make them especially prone to trigger AGNs through major galaxy mergers, although with only two such LSSs we are unable to determine their exact nature. 

While previous work has found mixed evidence for merger-induced AGN activity, this could be explained by multiple triggering scenarios. The differing results could be the result of differing sample selections, leading to many different cross-sections of the merger-induced AGN population. The data from the SG0023 \& SC1604 LSSs do provide evidence that galaxy mergers are important for triggering AGNs in some regimes, while the rest of our sample, and the similarities to our field AGNs, shows that at least one other mechanism plays a significant role. 

\section{Summary}

We studied AGNs and their hosts in 12 LSSs, using {\it Chandra} observations, optical and NIR imaging and spectroscopy, seeking to explore triggering mechanisms. We found a total of 61 AGNs, matched to spectroscopically confirmed hosts, within the LSSs in our sample. The hosts were associated with starbursts; the spectral features of their composite spectra indicated the AGN populations had less time since last starburst than their parent populations, on average. The hosts were also found to be associated with the green valley on a CMD, but did not seem to be associated with any particular environment within the LSSs. These results do not particularly favor any triggering mechanism. 

Two LSSs in our sample, SG0023 and SC1604, had a number of exceptional properties relative to the others. Their populations had disproportionately more X-ray luminous AGNs, they had the least average time since last starburst event, and both the AGN hosts and the overall galaxy populations had significantly higher close kinematic pair fractions. The last point indicates that galaxy mergers and interactions are happening more frequently in these LSSs compared to the others. The evidence suggests that the AGNs in SG0023 and SC1604 are being triggered by galaxy mergers. For the rest of the LSSs, other mechanisms should be dominant. This can be checked by looking at the field AGNs in our sample, which should have relatively low merger rates. Indeed, the field populations have low close kinematic pair fractions, and their X-ray luminosity distributions looked more like the other LSSs than like SG0023 and SC1604. The combination of these results suggests that major mergers are dominant triggers of AGNs in some regimes, such as those represented by SG0023 and SC1604, while other mechanisms dominate elsewhere and produce a population of lower luminosity AGNs. While there are many possibilities for other mechanisms, minor mergers are unlikely, as the same conditions that produce major mergers in SG0023 and SC1604 more frequently should produce minor ones as well. 

\bigskip

This material is based upon work supported by the National Aeronautics and Space Administration under NASA Grant Number NNX15AK92G. The authors thank Kirpal Nandra and Antonis Georgakakis for providing the Imperial reduction pipeline and their ongoing support of the software. The spectroscopic data presented herein were obtained at the W.M. Keck Observatory, which is operated as a scientific partnership among the California Institute of Technology, the University of California and the National Aeronautics and Space Administration. The Observatory was made possible by the generous financial support of the W.M. Keck Foundation. The authors wish to recognize and acknowledge the very significant cultural role and reverence that the summit of Mauna Kea has always had within the indigenous Hawaiian community.  We are most fortunate to have the opportunity to conduct observations from this mountain.

\appendix
\section{Completeness and Selection Effects}
\label{appendix}

\subsection{X-ray Completeness}
\label{sec:xraycomp}

Estimating the limiting luminosities of our {\it Chandra} observations is not straightforward. One of the main reasons for this is that, for each object, we had three chances to detect it, once in each of the three energy bands we considered: the soft (0.5-2.0 keV), hard (2.0-7.0 keV), and full (0.5-7.0 keV) bands. Note that the full band is simply the sum of the soft and hard bands, which adds the complication that the three bands are not independently observed. Furthermore, the relatively small number of counts per source brings us into the regime where Poisson statistics are highly relevant. So, to estimate the completeness of our X-ray detections, we carried out an MC simulation to take these effects, and others, into account. 

There are a number of variables that influence X-ray point source detection. Some are dependent on the field or observation, including the energy band of the observation, the background emission level and the exposure time. Others vary between sources or across the field of view, such the intrinsic luminosity of the sources, their redshifts, and their off-axis angle. The latter refers to the distance of the sky from the {\it Chandra} aimpoint, and the point spread function (PSF) is highly dependent upon it, varying from sub-arcsecond near the aimpoint to tens of arcseconds at the edges of the ACIS-I array (off-axis angles $\ga 10\arcsec$). Because of this, the X-ray completeness will vary substantially across the field of view. This makes describing the X-ray completeness difficult. If the entire field of view is considered for each LSS, the completeness will be quite low. However, we are not interested so much in finding every AGN in the field of view as comparing numerical measurements within our sample. As such, it is sufficient to examine a similar area within each observation, and compare completeness within those areas. We choose for this area a circle with radius 3$\arcmin$ centered on the {\it Chandra} aimpoint. 

For each LSS, we consider the completeness at specific redshifts, in the range $0.65<z<1.25$, with steps of $\Delta z=0.05$. For each redshift, we generated 100000 sources within the 3$\arcmin$ centered on the {\it Chandra} aimpoint for that observation. Each source was assigned a rest-frame, full-band X-ray luminosity drawn from a uniform distribution, in log-space, in the range 10$^{42}$ \ergs to 10$^{44.5}$ \ergs. To determine the luminosities in the soft and hard bands, each source was also assigned a hardness ratio, drawn from the smoothed distribution of hardness ratios we observed in our sample. 

After all points were generated, the observed fluxes were calculated using a power law spectral model with $\gamma = 1.4$. The next step is to convert these fluxes into expected photon counts observed. The factor for this conversion varies from source to source when calculated using the Imperial Pipeline discussed briefly in Section \ref{sec:datared} and in more detail in \citet{laird09} and \citet{nandra15}. However, it is approximately constant across each pointing. So, for each pointing in the soft and hard bands, we performed a linear fit to the relation between the net counts observed for each source and the flux calculated for that source. Using this value for all generated sources, we then estimated the expected photon counts in each band for each source. For each source, counts were `measured' within an aperture defined by the PSF at that position, designed to enclose 95\% of the flux for the soft band and 90\% in the hard band. The `measured' counts are the sum of the source counts and background counts, so the background counts also had to be simulated. To estimate the background levels, the images for each LSS were binned to 64 pixels (at $0\farcm492$ per {\it Chandra} pixel). For each source, the background level of the closest bin was used. The counts for each source were then drawn from a Poisson distribution defined by the total number of expected counts (source plus background) in the aperture. Background counts for each source were also generated for a background annulus, with inner radius and outer radius 1.5 and 2.5 times the aperture radii, respectively. These were drawn from a Poisson distribution defined by the expected number of background counts in the annulus. The net counts in the soft and hard counts could then be calculated by subtracting the total counts generated in the apertures by the scaled background counts generated in the annuli, which mimics our observations. The full-band net counts were then just the sum of the soft and hard band net counts.

\begin{figure*}
\includegraphics[width=\textwidth]{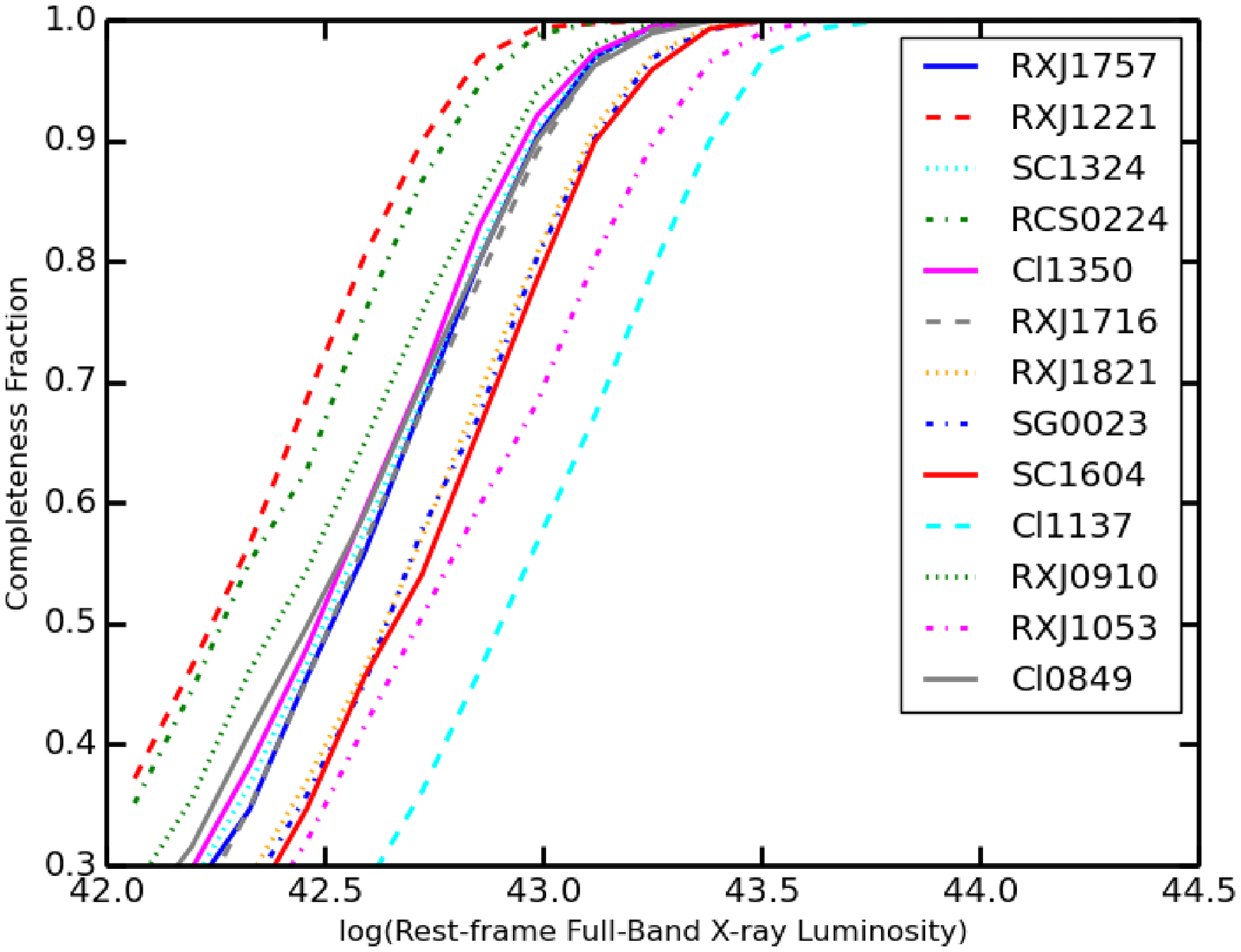}
\caption{
Results of X-ray completeness MC simulations are shown for each LSS as a function of rest-frame, full-band X-ray luminosity, at the redshift closest to $\langle z\rangle$ for that LSS. Completeness is measured here as the fraction of sources detected in a given luminosity bin in our simulation divided by the total number of simulated sources with that intrinsic luminosity. While Cl1137 is not included in our analysis, it is included in this figure for comparison. 
}
\label{LSSXraycompfig}
\end{figure*}

Each source could then be assigned a detection significance, using the formula $\sigma=C/\left(1+\sqrt{0.75+B}\right)$, where $C$ are the net photon counts and $B$ are the background counts in the aperture. As in our analysis, we define as detected all sources with $\sigma > 2$ in at least one band. We then bin all sources for each pointing and each redshift step into intrinsic rest-frame luminosity bins and calculate the completeness within each bin, defined here as the number of sources detected in that luminosity bin divided by the total number of sources in that luminosity bin. 

\begin{center}
\begin{table*}
\centering
\caption{Results of X-ray Completeness MC Simulation}
\begin{tabular}{lrrrrr}
\toprule
LSS Sub-sample &  \multicolumn{5}{c}{Percent Complete} \\
or Field &  \multicolumn{5}{c}{Luminosity Range (log)} \\
Population & 42-42.5 & 42.5-43 & 43-43.5 & 43.5-44 & 44.5-44.5\\
\midrule
Passive & 38.9 & 78.7 & 98.7 & \phantom{.}100 & 100\\
Intermediate & 32.8 & 74.0 & 98.7 & \phantom{.}100 & 100\\
SG0023 \& SC1604 & 24.4 & 61.6 & 95.1 & \phantom{.}100 & 100\\
High-$z$ & 33.8 & 70.0 & 95.4 & \phantom{.}100 & 100\\
Field (all) & 30.8 & 65.6 & 92.7& 99.7 & 100\\
Field ($z<0.8$) & 45.4 & 84.2 & 99.5& \phantom{.}100 & 100\\
Field ($0.8<z<0.96$) & 32.4 & 70.0 & 96.5& \phantom{.}100 & 100\\
Field ($z>0.96$) & 22.4 & 53.3 & 86.8& 99.4 & 100\\
\bottomrule
\label{comptab}
\end{tabular}
\end{table*}
\end{center}

The full-band completeness (hereafter, luminosities mentioned are full-band) is plotted for each LSS in Figure \ref{LSSXraycompfig}, at the redshift step closest to $\langle z\rangle$ (See Table \ref{strsumtab}) in each case. Note that Cl1137 is included in this plot, but not in the analysis in any other part of the paper. The reason for this is summarized in this figure, where the completeness for Cl1137 in 10-15\% lower than any other field across the luminosity range 10$^{42.5}$ \ergs to 10$^{43}$ \ergs. For all other fields, we can see completeness is almost unity above 10$^{43}$ \ergs. Completeness begins falling off below this value, and is at 35-70\% at 10$^{42.5}$ \ergs. The completeness fractions are more succinctly summarized in Table \ref{comptab}, where the fractions are averaged over the five luminosity bins between 10$^{42}$ \ergs and 10$^{44.5}$ \ergs\ in log space. In addition to the completeness fractions for each field, the average completeness fractions for each of the four sub-samples described in Section \ref{sec:SGC} are shown, as well as the combined completeness fractions for the field populations\footnote{Field galaxies are defined as those within the redshift range of the ORELSE sample ($0.65<z<1.28$), but outside the redshift bounds of the LSS in their {\it Chandra} observation.} across all observations\footnote{In these averages, the LSSs with two pointings, SC1324 and SC1604, are accorded double weighting. Additionally, Cl1137 is not included for these average measurements.}. We can see that the average completeness approximately ranges from 55\% to 90\% in the luminosity range 10$^{42.5}$ \ergs\ to 10$^{43}$ \ergs, but falls to approximately 25\% to 50\% below this for the LSSs. For this reason, we cut objects with luminosities below  10$^{42.5}$ \ergs. While the average completeness of each sub-sample is within 20\%, the histogram in Figure \ref{XrayLumfig} can be corrected to account for the individual completeness values. However, our results are not significantly affected by this correction. 

When considering the field galaxies, there is a wider range of completeness estimates for the 10$^{42.5}$ \ergs\ to 10$^{43}$ \ergs\ luminosity bin, ranging from 53\% for highest redshift galaxies to 84\% for the lowest galaxies. Correcting the field galaxy luminosity distribution in Figure \ref{XrayLumfig} accounts for some of the discrepancy between the different luminosity distributions of the three field galaxy redshift bins. 

\subsection{Spectroscopic Completeness}
\label{sec:photzcomp}

While some of our LSSs, such as SC1604, have nearly complete spectroscopy to $m_I\le24$, or $M_{red}\le-20.9$, this is not the case in general. Therefore, it is possible that some AGN hosts have been missed by our spectroscopic coverage so far. Since our initial spectral masks tended to target redder galaxies deemed more likely to be LSS members, varying levels of spectroscopic completeness could create biases between the observed AGN populations of the different LSSs. 

To determine the effects of incomplete spectroscopy, we used our photometric catalogs, described in Section \ref{sec:optobs}. For a subset of our sample, we used our photometry to estimate photometric redshifts, as discussed in Section \ref{sec:photoz} and in more detail in Tomczak et al. (2016, in prep.). In this process, we derived $P\left(z\right)$ for each source, which estimates the probability that any given redshift is the true redshift for that source. When $P\left(z\right)$ is integrated over a redshift range, it should give the probability that the source actually lies in that given range. So, we can use $P\left(z\right)$ to estimate the probability that sources without spectroscopic redshifts are members of a given LSS, by integrating over the redshift bounds of that LSS. With this method, we calculate the expected number of new AGNs that would be added as LSS members if we were to complete our spectroscopy, for the fields where photometric redshifts have been calculated. This is accomplished by calculating\begin{equation} 
\sum_i\int_{z_{LB}}^{z_{UB}} P_i\left(z\right) dz
\label{eq:Pz}
\end{equation}
where $z_{LB}$ and $z_{UB}$ are the redshift bounds of that LSS. The sum is over all sources without high quality spectroscopic redshifts, including only objects within our spectral coverage on the sky. 

This method relies on the sample being tested to have an approximately uniform true distribution in redshift space, at least in the area where $P\left(z\right)$ has a non-negligible overlap with the redshift range being considered. This can be seen by examining Equation \ref{eq:Pz}. Each term in the sum corresponds to one source, and adds between 0 and 1 to the total, which is the total number of expected new sources. Therefore, the sum must be less than or equal to the total number of sources being considered. This means that, if all these sources are true LSS members, Equation \ref{eq:Pz} necessarily underestimates the expected number of sources. The degree of underestimation is determined by the shape of the average $P\left(z\right)$ relative to the width of the LSS in redshift space. If the average $P\left(z\right)$ has substantial values outside of the redshift range, the estimated number of new sources will be considerably lower than the actual number. If the sample being considered includes sources actually located outside the LSS, those sources add to the sum in Equation \ref{eq:Pz}, which can compensate for the underestimation of the number of actual LSS members. As the true redshift distribution of the sample approaches uniformity, these outside sources bring the estimated number of sources in the LSS towards matching the true number, assuming $P\left(z\right)$ is similar in shape throughout the sample, on average. While we do not know the true redshift distribution of the sources being tested, this method should give us reasonable estimates of the number of sources we expect to add through completing spectroscopy.

When we apply this technique to the LSSs with photometric redshift catalogs, and ignoring objects where Equation \ref{eq:Pz} gives a probability of measurements less than 5\%, we estimate that 0.2, 1.7, 0, 0, 0.8, 0, and 0.2 X-ray point sources would be added as LSS members for SG0023, RXJ0910, SC1324, SC1604, RXJ1716, RXJ1757, and RXJ1821, respectively, with complete spectroscopy. On average over these LSSs, this is an increase in sample size of only $\sim$10\%, so our results should not be significantly affected by spectroscopic incompleteness. 
 
\bibliographystyle{mn2e}

\bibliography{rum}

\end{document}